%% file: main.tex
\documentclass[12pt]{article}
\usepackage{geometry}
\newgeometry{vmargin={15mm}, hmargin={17mm,17mm}}
\usepackage{titling}
\predate{}
\postdate{}

\normalsize
\usepackage{cite}
\usepackage{url}
\usepackage{amsmath}
\usepackage{array}
\usepackage{caption}
\usepackage{subcaption}

\usepackage{xcolor}

\usepackage{theorem}
\theoremheaderfont{\upshape\bfseries} \theoremstyle{plain}

\newtheorem{definition}{Definition}
\newtheorem{assumptions}{Assumption}

\newtheorem{remark}{Remark}

\newtheorem{theorem}{Theorem}

\newtheorem{lemma}{Lemma} 

\newcommand\numberthis{\addtocounter{equation}{1}\tag{\theequation}}

\usepackage{amssymb}
\usepackage{amsfonts}
\usepackage{graphicx}
\usepackage{psfrag}

\usepackage{hyperref}
\hypersetup{
    colorlinks=true,
    linkcolor=black,
    citecolor=black,
    filecolor=black,      
    urlcolor=blue,
    pdftitle={},
    pdfpagemode=FullScreen,
    }

\newcommand{\cC}{\mathcal{C}}
\newcommand{\cD}{\mathcal{D}}

\newcommand{\cF}{\mathcal{F}}
\newcommand{\cG}{\mathcal{G}}

\newcommand{\cN}{\mathcal{N}}

\newcommand{\cP}{\mathcal{P}}

\newcommand{\cS}{\mathcal{S}}

\newcommand{\cU}{\mathcal{U}}
\newcommand{\cV}{\mathcal{V}}
\newcommand{\cW}{\mathcal{W}}
\newcommand{\cX}{\mathcal{X}}
\newcommand{\cY}{\mathcal{Y}}
\newcommand{\cZ}{\mathcal{Z}}

\newcommand{\EE}{\mathbb{E}}

\newcommand{\NN}{\mathbb{N}}

\newcommand{\PP}{\mathbb{P}}

\newcommand{\RR}{\mathbb{R}}

\newcommand{\WW}{\mathbb{W}}
\newcommand{\XX}{\mathbb{X}}
\newcommand{\YY}{\mathbb{Y}}
\newcommand{\ZZ}{\mathbb{Z}}
\newcommand{\BS}{\mathbb{S}}
\newcommand*{\dd}{\, \mathrm{d}}

\newcommand{\vasti}{\bBigg@{3.5 }}
\newcommand{\vast}{\bBigg@{4}}
\newcommand{\Vast}{\bBigg@{5}}
\newcommand{\Vastt}{\bBigg@{7}}
\newcommand{\sC}{\mathsf{C}}
\newcommand{\sW}{\mathsf{W}}

\newcommand{\sh}{\mathsf{h}}
\newcommand{\sI}{\mathsf{I}}
\newcommand{\sX}{\mathsf{X}}
\newcommand{\sD}{\mathsf{D}}
\newcommand{\sP}{\mathsf{P}}

\newcommand{\DKL}{\mathsf{D}_{\mathsf{KL}}}
\newcommand{\Dn}{D_n}
\newcommand{\Bm}{B_m}
\newcommand{\Bphi}{B_m^\phi}

\usepackage{bbm}
\usepackage{comment}

\newcommand{\NDT}{h_\phi}
\newcommand{\NDTk}{h_{\phi,k}}

\newcommand{\NDTpush}{h_{\phi \sharp}}
\newcommand{\NDTkpush}{h_{\phi,k \sharp}}

\newcommand{\NE}{p^\psi_X}

\newcommand{\Dphi}{D_n^\phi}
\newcommand{\Dphik}{D_n^{\phi_k}}

\newcommand{\GNN}{\cG_\mathsf{nn}}
\newcommand{\GRNN}{\cG_\mathsf{rnn}}

\newcommand{\lebmeas}{\cP_{\mathsf{ac}}}
\newcommand{\mine}{\hat{\sI}_{{\mathsf{MI}}}}

\newcommand{\dine}{\hat{\sI}_{{\mathsf{DI}}}}

\newcommand{\argmax}{\mathop{\mathrm{argmax}}}
\renewcommand{\tilde}{\widetilde}
\renewcommand{\hat}{\widehat}
\newcommand{\indep}{\perp \!\!\! \perp}

\usepackage{lipsum}

\newcommand\blfootnote[1]{%
  \begingroup
  \renewcommand\thefootnote{}\footnote{#1}%
  \addtocounter{footnote}{-1}%
  \endgroup
}

\usepackage{setspace}
\usepackage{algorithm}
\usepackage[noend]{algpseudocode}
\newcommand{\algrule}[1][.1pt]{\par\vskip.1\baselineskip\hrule height #1\par\vskip.1\baselineskip}

\usepackage{tikz}
\usetikzlibrary{automata, positioning, calc}
\usetikzlibrary{arrows.meta, positioning, calc}

\tikzstyle{block_D} = [rectangle, rounded corners, minimum width=2.5cm, minimum height=1.4cm, align=center, draw=black, fill= black!5]

\tikzstyle{block_E} = [rectangle, rounded corners, minimum width=0.8cm, minimum height=0.8cm, align=center, draw=black, fill= black!5]

\tikzstyle{block_C} = [rectangle, minimum width=1cm, minimum height=0.01cm, draw=white, fill=white]
 
\tikzstyle{small_b} = [rectangle, minimum width=0.000001cm, minimum height=0.000001cm, draw=none, fill=black!20]

\tikzstyle{small_block} = [rectangle, minimum width=1.3cm, minimum height=1.3cm, draw=black, fill=white, rounded corners]

\tikzstyle{lstm_block} = [rectangle, rounded corners, minimum width=4cm, minimum height=4cm, align=center, draw=black, fill= black!4]

\tikzstyle{inv1_block} = [rectangle, minimum width=0.000001cm, minimum height=0.000001cm, draw=none, fill=none]

\begin{document}
\raggedbottom
%
\title{Neural Estimation and Optimization of Directed Information over Continuous Spaces}
%
%
%

\author{Dor~Tsur,
        Ziv~Aharoni,
        Ziv~Goldfeld,
        and~Haim~Permuter}
\date{}

\markboth{}%
{}
%

\maketitle

\begin{abstract}
This work develops a new method for estimating and optimizing the directed information rate between two jointly stationary and ergodic stochastic processes. Building upon recent advances in machine learning, we propose a recurrent neural network (RNN)-based estimator which is optimized via gradient ascent over the RNN parameters.
The estimator does not require prior knowledge of the underlying joint and marginal distributions.
The estimator is also readily optimized over continuous input processes realized by a deep generative model. 
We prove consistency of the proposed estimation and optimization methods and combine them to obtain end-to-end performance guarantees. Applications for channel capacity estimation of continuous channels with memory are explored, and empirical results demonstrating the scalability and accuracy of our method are provided.
When the channel is memoryless, we investigate the mapping learned by the optimized input generator.
\end{abstract}


%


\section{Introduction}
\blfootnote{Part of this work was presented at the International Symposium on Information Theory (ISIT) 2020 \cite{aharoni2020capacity}}
Directed information (DI), introduced by Massey \cite{massey1990causality}, quantifies the amount of information one stochastic process causally conveys about another. It possesses structural properties that render it as the natural causal analog of mutual information (MI), and it emerges as the solution to various operational problems involving causality \cite{raginsky2011directed}.
Applications of DI are abundant, from the capacity of communication channels with or without memory, which is generally given by maximized DI \cite{gallager1968information,permuter2009finite}, to causal hypothesis testing and portfolio theory  \cite{permuter2011interpretations}, where DI intricately relates to optimal tests and investment strategies, respectively. 
DI has also seen a myriad applications in machine learning \cite{battiti1994using,bell1995information,higgins2016beta,shwartz2017opening,goldfeld2018estimating}, neuroscience \cite{dimitrov2011information,quinn2011estimating,wibral2014directed}, and control
\cite{touchette2000information, grocholsky2002information}, to name a few.
It is oftentimes of interest not only to evaluate DI but also to optimize it (e.g., to characterize capacity, to bound growth rates of optimal portfolios, to extract informative features, etc.). 
However, this optimization is challenging since analytic computation of DI requires knowledge of the underlying probability law, which is typically unavailable in practice. Furthermore, even when the probability law is given, tractable DI characterizations that lend well for optimization are rare \cite{boche2020shannon, grigorescu2022capacity}, as it is generally given by a multiletter expression.
To address this, the goal of the paper is to develop a computable and provably accurate estimate of DI.

\subsection{Estimation and Optimization of Directed Information}
Existing estimators of DI operate under rather restrictive assumptions on the data, hence covering a small class of problems. DI estimation between discrete-valued processes using universal probability assignments and context tree weighting was studied in 
\cite{jiao2013universal}. Their estimator is provably consistent, but requires that the depth of the context tree is greater than the assumed memory of the processes.
An approach based on maximum likelihood estimation of the associated PMF was developed in \cite{quinn2015directed}. 
However, both \cite{jiao2013universal} and \cite{quinn2015directed} are limited to the class of discrete-valued, stationary Markov processes of relatively small order. Continuous-valued processes, which are of central practical interest, were treated in \cite{murin2017k,rahimzamani2018estimators} using $k$ nearest neighbors ($k$NN) estimation techniques, but as the memory or dimension of the data increase, the performance of $k$NN-based techniques deteriorates, due to the curse of dimensionality \cite{marimont1979nearest}.

Neural estimation is a modern technique for estimating divergences and information measures. Originally proposed in \cite{belghazi2018mutual}, the MI neural estimator (MINE) parametrizes the Donsker-Varadhan (DV) variational form \cite{donsker1983asymptotic} by a neural network (NN), and optimizes it over a parameter space.
Several variations of the MINE were proposed in followup work, e.g., replacing the DV representation with other variational lower bounds \cite{poole2018variational,song2019understanding}, or by incorporating auxiliary distributions 
\cite{chan2019neural}. Consistency of MINE in the infinite-width NN regime was established in \cite{belghazi2018mutual}, and non-asymptotic error bounds were later derived in \cite{sreekumar2021non,sreekumar2021neural}. The latter, in particular, showed that MINE is minimax optimal under appropriate regularity assumptions on the distributions (see also \cite{mcallester2020formal} for formal limitations on MINE performance).
For data with memory, \cite{zhang2019itene} leveraged MINE for transfer entropy, while \cite{molavipour2021neural} constructed a conditional MI estimator and extended it to DI between 1st order Markov processes.

In many applications, it is of interest to optimize DI over the involved processes. A prominent example is channel capacity computation, which is also the main application considered herein. Tools from dynamic programming were used in \cite{permuter2008capacity,elishco2014capacity} to estimate the feedback capacities of a class of binary finite state channels (FSCs). This approach was later generalized to large discrete alphabets using reinforcement learning \cite{aharoni2020reinforcement}. Another approach towards maximizing information measures relies on the Blahut-Arimoto (BA) algorithm \cite{blahut1972computation,arimoto1972algorithm}, originally proposed for MI maximization between discrete random variables. Subsequently, the algorithm was extended to FSCs \cite{vontobel2008generalization}, to DI \cite{naiss2012extension}, and to MI between continuous random variables \cite{dauwels2005numerical}. The main drawback of BA algorithms is that they require full knowledge of the involved densities or the availability of consistent estimates thereof. Moreover, the continuous BA algorithm is based on space quantization, and therefore its computational complexity grows exponentially with the variables dimension.

\subsection{Contributions}

Building on the computational potency of modern machine learning techniques, we develop herein a neural estimation and optimization framework for the DI rate between continuous-valued stochastic processes. Inspired by \cite{belghazi2018mutual,chan2019neural}, we derive the DI neural estimator (DINE) by expressing DI in terms of certain Kullback-Leibler (KL) divergences (plus cross-entropy residuals) and invoking the DV representation to arrive at a variational form. To account for causal dependencies, we parametrize the DV feasible set with the set of recurrent neural networks (RNNs) and approximate expected values by sample means. This results in a parametrized empirical objective that lends well to gradient-based optimization. We prove that the DINE is consistent whenever the stochastic processes are stationary and ergodic. The proof is based in a generalized version of Birkhoff's ergodic theorem \cite{breiman1957individual}, martingale analysis, and the universal approximation property of RNNs~\cite{schafer2006recurrent}.

Having the DINE, we consider optimization of the estimated DI rate over the input stochastic process.
To that end, we simulate the input process by an RNN deep generative model, whose parameters can be tuned to increase the estimated DI rate. By jointly optimizing the DINE and the input generative model, we obtain an estimation-optimization scheme for estimating the capacity of continuous channels with memory. Consistency of the overall method is established using the functional representation lemma (FRL) \cite{el2011network,li2018strong} and universal approximation arguments \cite{schafer2006recurrent}. We provide an extensive empirical study of the proposed method, demonstrating its efficiency and accuracy in estimating the feedforward and feedback capacities of various channels with and without memory, encompassing the average and peak power constrained additive white Gaussian noise channels (AWGN) \cite{ozarow1990capacity,raginsky2008information,thangaraj2017capacity}, moving-average (MA) AWGN \cite{yang2007feedback} and MIMO auto-regressive (AR) AWGN channels \cite{sabag2021feedback}. 
Lastly, we discuss the structure of the learned optimal input distribution and furnish connections to probability integral transforms.
We note that following the earlier conference version of this paper \cite{aharoni2020capacity}, several neural optimization techniques were proposed \cite{ mirkarimi2021neuralcap,letizia2021capacity, letizia2021discriminative} and an empirical comparison was the focus of \cite{mirkarimi2022perspective}.
However, these methods are only applicable to memoryless channels.


\subsection{Organization}
The text is organized as follows. Section \ref{sec:prel} provides preliminaries and technical background.
Section \ref{sec:main_results} summarizes the main results of this paper.
Section \ref{sec:dine} derives the DINE, provides theoretical guarantees, and discusses its implementation.
The optimization procedure of DINE over continuous-valued input processes is the focus of Section \ref{sec:cont_opt}, where consistency of the overall method and implementation details are also given. Section \ref{sec:cont_opt_exp} provides empirical results for channel capacity estimation. Proofs are given in Section \ref{sec:proofs}, while Section \ref{sec:conclusion} provides concluding remarks and discusses future research directions.

\section{Background and preliminaries}\label{sec:prel}

\subsection{Notation}
Subsets of the $d$-dimensional Euclidean space are denoted by calligraphic letters, e.g., $\cX\subseteq \RR^d$.
For any $n\in\NN$, $\cX^n$ is the $n$-fold Cartesian product of $\cX$, while $x^n=(x_1,\dots,x_n)$ denotes an element thereof.
For $i,j\in\ZZ$ with $i\leq j$, we use the shorthand $x_i^j:=(x_i,\dots,x_j)$; the subscript is omitted when $i=1$.
We denote by $(\Omega,\cF,\PP)$ the underlying probability space on which all random variables are defined, with $\EE$ denoting expectation.
The set of all Borel probability measures on $\cX\subseteq\RR^d$ is denoted by $\cP(\cX)$. The subset of $\cP(\cX)$ of Lebesgue absolutely continuous measures is denoted by $\lebmeas(\cX)$. The density of $P\in\lebmeas(\cX)$ is designated by its lowercase version $p$; $n$-fold product extensions of $P$ and $p$ are denoted by $P^{\otimes n}$ and $p^{\otimes n}$, respectively. 
Random variables are denoted by upper-case letters, e.g., $X$, using the same conventions as above for random vectors.
Stochastic processes are denoted by blackboard bold letters, e.g., $\XX:=(X_i)_{i\in\NN}$.


For $P,Q\in\cP(\cX)$ such that $Q\ll P$, i.e., $Q$ is absolutely continuous with respect to (w.r.t.) $P$, we denote the Radon-Nykodim derivative of $P$ w.r.t. $Q$ by $\frac{\mathrm{d}P}{\mathrm{d}Q}$.
The KL divergence between $P$ and $Q$ is
$\DKL(P\|Q):=\EE_P\big[\log\frac{\mathrm{d}P}{\mathrm{d}Q}\big]$. If $Q\in\lebmeas(\cX)$ with probability density function (PDF) $q$, then the cross-entropy between $P$ and $Q$ is
$\sh_{\mathsf{CE}}(P,Q):=-\EE_P\left[\log q\right]$.
The MI between $(X,Y)\sim P_{XY}\in\cP(\cX\times \cY)$ is $\sI(X;Y) := \DKL(P_{XY}\|P_X\otimes P_Y)$, where $P_X$ and $P_Y$ are the marginals of $P_{XY}$. The differential entropy of $X\sim P$ is $\sh(X) := \sh_{\mathsf{CE}}(P,P)$ whenever $P\in\lebmeas(\cX)$.
We denote the convolution between two probability measures $\mu$ and $\nu$ with $(\mu*\nu)(A):=\int\int\mathbbm{1}_A(x+y)\dd\mu(x)\dd\nu(y)$ and $\mathbbm{1}_A$ as the indicator of $A$
For an open set $\cU\subseteq\RR^d$ and $k\in\NN$, the class of functions such that all partial derivatives up to order $k$ exist and are continuous is denoted by $\cC^k(\cU)$, with $\cC(\cU) := \cC^0(\cU)$ and we denote by $\partial^j_{x_i}f$ the $j$th order partial derivative of $f$ w.r.t. $x_i$.

\subsection{Directed Information and Channel Capacity}\label{subsec:di}
Originally proposed by Massey \cite{massey1990causality}, DI quantifies the amount of information one sequence of random variables causally conveys about another.
\begin{definition}[Directed information]
Let $(X^n,Y^n)\sim P_{X^n Y^n}\in\cP(\cX^n\times\cY^n)$.
The DI from $X^n$ to $Y^n$ is 
\begin{equation}
    \sI(X^n\to Y^n):= \sum_{i=1}^n \sI(X^i;Y_i|Y^{i-1}).
\end{equation}
\end{definition}
DI entails the concept of causal conditioning, i.e., conditioning only on present and past values of the sequences, which is seen through its decomposition using causal conditioned (CC) entropies \cite{kramer1998directed}.
For $(X^n,Y^n)\sim P_{X^n Y^n}\in\cP(\cX^n\times\cY^n)$, the entropy of $Y^n$ CC on $X^n$ is given by
$$
    \sh\left(Y^n \| X^n\right):=  \mathbb{E}\left[ -\log p_{Y^n\|X^n}\left(Y^n\|X^n\right) \right],
$$
where $p_{Y^n\|X^n}\left(y^n \| x^n\right):= \prod_{i=1}^n p_{Y_i|Y^{i-1},X^i}\left(y_i|y^{i-1},x^i\right)$ is the CC-PDF of $Y^n$ given $X^n=x^n$.
As the CC entropy can be expressed as $\sh(Y^n\|X^n):=\sum_{i=1}^n \sh(Y^i|X^i,Y^{i-1})$, we have the following representation for DI:
\begin{equation}\label{eq:di_ent_diff}
   \sI\left(X^n\to Y^n\right)=\sh\left(Y^n\right)-\sh\left(Y^n\|X^n\right).
\end{equation}
This poses DI as the reduction in the uncertainty about $Y^n$ as a result of causally observing (the elements of) $X^n$.
Since DI (as well as MI) tends to grow with the number of observations, the appropriate figure of merit when considering stochastic processes is the DI \emph{rate}.
\begin{definition}[Directed information rate]
Let $\XX$ and $\YY$ be jointly stationary stochastic processes. The DI rate from $\XX$ to $\YY$ is given by
\begin{equation}
    \sI(\XX\to\YY):= \lim_{n\to\infty}\frac{1}{n}\sI(X^n\to Y^n).
\end{equation}
\end{definition}
The limit exists whenever the processes are jointly stationary \cite{CovThom06}.
Due to the averaging, the DI rate 
captures prominent interactions, while the effect of transient phenomena decays to zero.

\begin{remark}[Channel capacity]
We consider channels with and without a feedback
link from the channel output back to the encoder.
The feedforward capacity of a sequence of channels $\{P_{Y^n\|X^n}\}_{n\in\NN}$ is
\cite{gallager1968information}\footnote{This formula assumes the so-called information stability property (see \cite{dobrushin1963general}).}
\begin{equation}
    C_{\mathsf{FF}} = \lim_{n \rightarrow \infty}  \sup_{P_{X^n}}{\frac{1}{n} \sI(X^n;Y^n)}.\label{eq:CFF}
\end{equation}
In the presence of feedback, the capacity becomes \cite{kim2008coding}
\begin{equation}
    C_{\mathsf{FB}} = \lim_{n \rightarrow \infty}  \sup_{P_{X^n \| Y^{n-1}}}{\frac{1}{n} \sI(X^n \rightarrow Y^n)}\label{eq:CFB}.
\end{equation}
The achievability of \eqref{eq:CFF} and \eqref{eq:CFB} is discussed in \cite{dobrushin1963general} and \cite{kim2008coding}, respectively.
As shown in \cite[Theorem 1]{massey1990causality}, when feedback is not present, the optimization problem \eqref{eq:CFB} (which amounts to optimizing over $P_{X^n}$ rather than $P_{X^n\|Y^n}$) coincides with \eqref{eq:CFF}.
Thus, DI provides a unified framework for the calculation of both feedforward and feedback capacities. 
\end{remark}

\subsection{Neural Networks and Recurrent Neural Networks}\label{subsec:prel_mine}
The class of shallow NNs with fixed input and output dimensions is defined as follows \cite{hornik1989multilayer}. 
\begin{definition}[NN function class]\label{def:NN_function_class}
For the ReLU activation function
$\sigma_\mathsf{R}(x) = \max(x,0)$ and $d_{\mathsf{i}},d_{\mathsf{o}} \in\NN$, define the class of neural networks with $k\in\NN$ neurons as:
\begin{equation}
    \cG_k^{(d_{\mathsf{i}},d_{\mathsf{o}})}:=\left\{g:\RR^{d_{\mathsf{i}}}\to\RR^{d_{\mathsf{o}}}: g(x)=\sum_{j=1}^k\beta_j \sigma_\mathsf{R}( \mathrm{W}_j x-b_j),\ x\in\RR^{d_{\mathsf{i}}} \right\},\label{eq:NN_def}
\end{equation}
where $\sigma_\mathsf{R}$ acts component-wise, $\beta_j\in\RR, \mathrm{W}_{j}\in \RR^{d_{\mathsf{o}} \times d_{\mathsf{i}}}$ and $ b_j\in\RR^{d_{\mathsf{o}}}$ are the parameters of $g\in \cG_k^{(d_{\mathsf{i}},d_{\mathsf{o}})}$. 
Then, the class of NNs with input and output dimensions $(d_{\mathsf{i}},d_{\mathsf{o}})$ is given by
\begin{equation}
    \GNN^{(d_{\mathsf{i}},d_{\mathsf{o}})} := \bigcup_{k\in\NN} \cG_k^{(d_{\mathsf{i}},d_{\mathsf{o}})}.\label{eq:grnn_union}
\end{equation}
\end{definition}
NNs form a universal approximation class under mild smoothness conditions \cite{hornik1989multilayer}. However, feedforward networks such as those defined in \eqref{eq:NN_def} cannot capture temporal evolution, which is inherent to DI. Therefore, our neural estimator employs RNNs \cite{jin1995universal}, as defined next.

\begin{definition}[RNN function class]\label{def:RNN_function_class}
Let $t= 1,\dots,T$, $\alpha\in(-1,1)$, $a\in\RR^{k}$, $\mathrm{B}\in\RR^{d_{\mathsf{i}}\times k}$ and $\mathrm{C}\in\RR^{k\times d_{\mathsf{o}}}$.
The class $\GRNN^{(d_{\mathsf{i}},d_{\mathsf{o}},k)}$ of RNNs with $k$ neurons is the set of nonlinear systems with the following structure:
\begin{align*}
    s_{t+1} &= -\alpha x_{t} + a\sigma_\mathsf{S}(s_{t}+\mathrm{B}u_{t}), \quad u_t\in\RR^{d_{\mathsf{i}}},s_t\in\RR^{k}\\
    x_t &= \mathrm{C} s_{t}, \hspace{3.82cm} x_t\in\RR^{d_{\mathsf{o}}},
\end{align*}
where the sigmoid activation, denoted $\sigma_\mathsf{S}(x)=(1+\exp(-x))^{-1}$, acts component-wise. The class of RNNs with dimensions $(d_{\mathsf{i}},d_{\mathsf{o}})$ is defined as
\begin{equation}\label{definition:RNN_class}
    \GRNN^{(d_{\mathsf{i}},d_{\mathsf{o}})}:= \bigcup_{k\in\NN}\GRNN^{(d_{\mathsf{i}},d_{\mathsf{o}},k)}.
\end{equation}
\end{definition}
Note that both $\cG_k^{(d_{\mathsf{i}},d_{\mathsf{o}})}$ and $\GRNN^{(d_{\mathsf{i}},d_{\mathsf{o}},k)}$ are parametric models whose (finitely many) parameters belong to some parameter space $\Theta\subset\RR^d$, for an appropriate dimension $d$. When $k$ is fixed, interchangeably denote functions from the above classes explicitly, as $g\in\cG_k^{(d_{\mathsf{i}},d_{\mathsf{o}})}$, or in their corresponding parametrized form: $g_\theta$ where $\theta\in\Theta$.

\subsection{Mutual Information Neural Estimation}
The mutual information neural estimator (MINE) \cite{belghazi2018mutual} is a NN-based estimator of the MI between two random variables. The technique relies on the DV variational representation of KL divergence \cite[Theorem 3.2]{donsker1983asymptotic}.

\begin{theorem}[DV representation]\label{theorem:DV}
For any $P, Q\in\cP(\cX)$, we have

\begin{equation}
    \DKL\left(P \middle\| Q\right) = \sup_{f: \cX \to \mathbb{R}}\mathbb{E}_P\left[ f \right] -\log\left(\mathbb{E}_Q[ e^{f} ]\right)\label{eq:DV},
\end{equation}
where the supremum is taken over all measurable functions $f$ for which expectations are finite. 
\end{theorem}
Given $n$ pairwise independent and identically distributed (i.i.d.) samples $\Dn:=(X^n,Y^n)$ from $P_{XY}\in\cP(\cX\times\cY)$, the MINE parametrizes $f$ by a NN $g\in\GNN:=\GNN^{(d_x+d_y,1)}$
and approximates expectations by sample means:

\begin{equation}
    \mine(\Dn) := \sup_{g\in\GNN}\underbrace{\frac{1}{n}\sum_{i=1}^n g(X_i,Y_i) - \log\left( \frac{1}{n}\sum_{i=1}^n  e^{g(X_i,\bar{Y_i})}\right)}_{=\mine(\Dn,g)}\label{eq:mine_objective},
\end{equation}
where 
$(X_i,\bar{Y}_i)\sim P_X\otimes P_Y$.
The functions over which we optimize the DV objective are termed DV potentials.
We stress that only the correlated samples from  $\Dn$ are given, so negative (i.e., independent) samples must be constructed from them, e.g., by random permutation \cite{belghazi2018mutual}.
In \cite[Theorem 2]{belghazi2018mutual} the strong consistency of MINE is proved, i.e., $\lim_{n\to\infty}\mine(\Dn)=\sI(X;Y)$, $\PP$-almost surely (a.s.).
\begin{remark}[Non-asymptotic neural estimation error bound]
Non-asymptotic error bounds for neural estimation of $f$-divergence were recently derived in \cite{sreekumar2021neural}.
Specifically, they established bounds on the effective (approximation plus empirical estimation) error of a neural estimator realized by a $k$-neuron shallow NN with bounded parameters and $n$ data samples. Instantiating their result for the $\DKL(P\|Q)$ with $P=P_{XY}$ and $Q=P_X\otimes P_Y$ yields an
$O\big(d^{1/2}k^{-1/2}+ d^{3/2}(\log k)^{7}n^{-1/2}\big)$
error bound for MI estimation, uniformly over a class of sufficiently regular $d$-dimensional distributions with bounded supports. Evidently, there is a fundamental tradeoff between the two sources of error: while good approximation needs the NN class to be rich and expressive, empirical estimation error bounds rely on controlling complexity.
\end{remark}

Due to the consistency of the MINE, and since parameterization can only shrink the DV function class, it provably lower bounds the ground truth MI in the limit of large samples. 
\begin{lemma}[MINE lower bounds MI]\label{lemma:mine_lb}
For any $g\in\GNN$, we have 
\begin{equation}
    \sI(X;Y) \geq \lim_{n\to\infty}\mine(\Dn,g), \qquad \PP-a.s.
\end{equation}
\end{lemma} 
This property implies that the probability that MINE will overestimate MI is small.
This property is central when the target MI is the underlying capacity of some communication channel, as we can state that the estimate provides a lower bound of it at worst.
This property will be further discussed in the context of the proposed methods.

\section{Main Results}\label{sec:main_results}

This work develops a principled framework for neural estimation and optimization of information measures, which is then leveraged to estimate the feedforward and feedback capacities of general channels. To that end we propose the DINE, which generalizes the MINE for DI rate, and develop methods for optimizing MINE and DINE over continuous channel input distributions. 
While channel capacity estimation is the focus of this work, the proposed estimation and optimization techniques are applicable to any DI optimization scenario.

\subsection{Directed Information Neural Estimation}\label{subsec:main_results_dine}
We set up the DINE, state its consistency, and provide a pseudo-algorithm for its computation. 
We construct the DINE as the difference between two DV-based KL estimators.
Given a sample $\Dn=(X^n,Y^n)\sim P_{X^nY^n}$ and RNNs $g_y\in\GRNN^{Y} := \GRNN^{(d_y,1)}$ and $g_{xy}\in\GRNN^{XY} := \GRNN^{(d_y+d_x,1)}$, the DINE is given by
$$
\dine(\Dn)
    :=\sup_{g_{xy}\in\GRNN^{XY}}\hat{\sD}_{Y\|X}(\Dn,g_{xy})-\sup_{g_y\in\GRNN^{Y}}\hat{\sD}_{Y}(\Dn,g_y),
$$
where $\hat{\sD}_{Y}, \hat{\sD}_{Y\|X}$ are given by
\begin{subequations}
\begin{align}
   \hat{\sD}_Y(\Dn, g_y) &:= \frac{1}{n}\sum_{i=1}^n{g_y}\left(Y^i\right)-\log\left(\frac{1}{n}\sum_{i=1}^n e^{g_y\left(\widetilde{Y}_i,Y^{i-1}\right)}\right)\\
   \hat{\sD}_{Y\|X}(\Dn,g_{xy}) &:= \frac{1}{n}\sum_{i=1}^n{g_{xy}}\left(Y^i,X^i\right)-\log\left(\frac{1}{n}\sum_{i=1}^n e^{g_{xy}\left(\widetilde{Y}_i,Y^{i-1},X^i\right)}\right),
\end{align}\label{eq:DINE_KL_est_main}%
\end{subequations}
and $\tilde{Y}^n\stackrel{i.i.d.}{\sim}\mathsf{Unif}(\cY)$.
A full derivation of the estimator and further implementation details are provided in Section \ref{sec:dine}.
As stated next, the DINE is a consistent estimator of the DI rate. 
\begin{theorem}[Consistency]\label{theorem:DINE_consistency}
Suppose $\XX$ and $\YY$ are jointly stationary ergodic stochastic processes. Then the DINE is a strongly consistent estimator of $\sI(\XX\to\YY)$, i.e., for every $\epsilon>0$ there exists $N\in\NN$ such that for every $n>N$ we have
\begin{equation}
    \Big|\hspace{1mm}\dine(\Dn) - \sI(\XX\to\YY) \Big| \leq \epsilon, \qquad \PP-a.s.\label{eq:consistency_error}
\end{equation}
\end{theorem}
To compute the DINE in practice notice that $\GRNN^Y$ and $\GRNN^{XY}$ are parametric classes.
We fix $k$ and take their $k$-dimensional counterparts whose (finitely many) parameters belong to some parameter space $\Theta\subset\RR^d$, for an appropriate dimension $d$.
We therefore denote the DINE RNNs with $g_{\theta_y}$ and $g_{\theta_{xy}}$, and optimize the DINE objective over their parameters $(\theta_y,\theta_{xy})$ via stochastic gradient-ascent, as delineated in Algorithm \ref{alg:DINE} below.

\begin{algorithm}[h]
\setstretch{1.15}
    \caption{DINE}
    \label{alg:DINE}
    \textbf{Input:} Dataset $\Dn$.\\
    \textbf{Output:} $\dine(\Dn)$ DI rate estimate.
    \algrule
    \begin{algorithmic}
    \State Initialize $g_{\theta_y}$, $g_{\theta_{xy}}$ with parameters $\theta_{y} ,\theta_{xy}$.
    \State \textbf{Step 1 -- Parameter optimization:}
    \Repeat
    \State\hspace{-3mm}Draw a batch $\Bm$ for $m<n$ \& sample $P_{\tilde{Y}}$.
    \State\hspace{-3mm}Compute $\hat{\sD}_{Y \| X}(\Bm, g_{\theta_{xy}})$, $\hat{\sD}_{Y}(\Bm, g_{\theta_{y}})$ using \eqref{eq:DINE_KL_est_main}.
    
    \State\hspace{-3mm}Update networks parameters:\\
    \quad\quad\quad$\theta_{xy} \leftarrow \theta_{xy} + \nabla_{\theta_{xy}}\hat{\sD}_{Y \| X}(\Bm, g_{\theta_{xy}})$\\
    \quad\quad\quad$\theta_{y} \leftarrow \theta_{y} +\nabla_{\theta_{y}}\hat{\sD}_{Y}(\Bm, g_{\theta_{y}})$
    \Until{convergence.}
    
    \State \textbf{Step 2 -- Evaluation:} Evaluate over a sample $\Dn$ and subtract losses to obtain $\dine(\Dn)$ \eqref{eq:dine_objective}.
    \end{algorithmic}
\end{algorithm}

\subsection{DINE Optimization over Continuous Spaces}\label{sec:main_results_cont}

Given a sequence of transition kernels $\{P_{Y^n\|X^n}\}_{n\in\NN}$ that models a communication channel, we propose a method for optimizing the DINE over continuous input distributions.
Specifically, we employ an RNN generative model termed the \textit{neural distribution transformer} (NDT),
denoted $\NDT\in\GRNN^{(d_x,d_x,k)}$, where $k,d_x\in\NN$ and $\phi\in\Phi$ are its parameters.
Let $U^n\sim P^{\otimes n}_U$ for some $P_U\in\cP_{\mathsf{ac}}(\cU)$ and $\cU\subset\RR^{d_x}$. We define $\NDT$ through the following recursive relation 
$$
\NDT: (U_i,Z^\phi_{i-1}) \mapsto X^\phi_i, \qquad i=1,\dots,n,
$$
where $Z^\phi_i$ is determined according to whether feedback is present or not.
By sampling $P_U$ and passing those samples through $\NDT$ and the channel, we generate a dataset $\Dphi=(X^{\phi,n},Y^{\phi,n})$.
We optimize the DINE over $\phi$ such that $\Dphi$ corresponds to the distribution that maximizes the DINE. 
The optimization is executed via stochastic gradient ascent.

We prove the convergence of the joint estimation-optimization method.
We assume that both the channel and input process adhere to a recursive nonlinear and stationary state space model.
We further assume that the channel output and state mappings, given by $f_\mathsf{y}$ and $f_{\mathsf{z}}$, meet some Lipschitz continuity criterion (this is summarized by Assumption \ref{assumption:dine_ndt_channel_assumption}, in Section \ref{subsec:dine_ndt_consistency}).
We denote the class of such input processes by $\sX_{\cS}$ and denote the maximal DI rate over $\sX_{\cS}$ by $\underline{\sC}_s$ .
We propose the following
\begin{theorem}[Strong consistency of the DINE-NDT method]\label{theorem:ndt_consistency0}
Fix $\epsilon>0$, let $U^n\sim P^{\otimes n}_U$, and consider the continuous unifilar state channel $\{P_{Y_i|Y^{i-1},X^i}\}_{i\in\NN}$, where $f_\mathsf{y}, f_\mathsf{z}$ satisfy Assumption \ref{assumption:dine_ndt_channel_assumption}.
Then there exists $N\in\NN$ such that for every $n>N$, we have
\begin{equation}
    \left|\underline{\sC}_s-\dine^\star(U^n)\right|\leq\epsilon,\qquad \PP-a.s.,
\end{equation}
where $\dine^\star(U^n)=\sup_{\NDT\in\GRNN^{(d_x,d_x)}}\dine(\Dphi,\NDT)$.
\end{theorem}

\begin{algorithm}[t]
\caption{Continuous DINE optimization}
\label{alg:cap_est}
\setstretch{1.15}
\textbf{Input:} Continuous channel, feedback indicator.\\
\textbf{Output:} $\dine^\star(U^n)$, optimized NDT.
\algrule
\begin{algorithmic}
\State Initialize $g_{\theta_y}, g_{\theta_{xy}}$ and $\NDT$ with parameters $\theta_{y}, \theta_{xy}, \phi$.
    \If{feedback indicator}
        \State Add feedback to NDT.
    \EndIf
\Repeat
\State Draw noise $U^m$, $m<n$. 
\State Compute $\Bphi$ using NDT and channel
\If{training DINE}
\State Perform DINE optimization according to step 1 in Algorithm \ref{alg:DINE}.
\Else \hspace{0.15cm}(Train NDT)
\State Compute $\dine(\Bphi, g_{\theta_{y}},g_{\theta_{xy}},\NDT)$ using \eqref{eq:DINE_KL_est_main}. 
\State Update NDT parameters: \\
\hspace{\algorithmicindent}\hspace{\algorithmicindent}$\phi \leftarrow \phi + \nabla_{\phi}\dine(\Bphi, g_{\theta_{y}} ,g_{\theta_{xy}},\NDT)$
\EndIf
\Until{convergence.}
\State Draw $U^n$ and perform a Monte Carlo evaluation of  $\dine(D^\phi_n)$. \\
\Return $\dine^\star(U^n)$, optimized NDT.
\end{algorithmic}
\end{algorithm}

For memoryless channels, where capacity is given by the maximized MI, we consider MINE optimization and identify the optimized NDT structure via multivariate generalization of the capacity achieving input cumulative distribution function (CDF), obtained by vectorizing the product of conditional CDFs of the entries of $X$ (see Theorem \ref{theorem:opt_ndt}).
For the full details of the theoretical guarantees, see Section \ref{subsec:ndt_theo}.

As described in Algorithm \ref{alg:cap_est}, the joint DI estimation-maximization procedure involves alternating optimization between the DINE and NDT models.
In Section \ref{sec:cont_opt_exp}, we demonstrate the end-to-end procedure by estimating the capacity of several channels, with and without memory, accounting for both feedforward and feedback capacities.
We empirically demonstrate the accuracy of the algorithm by comparing it with known results/bounds and analyse the optimized NDT model.

\section{Directed Information Neural Estimation}\label{sec:dine}
This section describes the DI estimation method introduced in Section \ref{subsec:main_results_dine}. 
We consider two jointly stationary and ergodic processes $\XX$ and $\YY$, supported on $\cX\subseteq\RR^{d_x}$ and $\cY\subseteq\RR^{d_y}$, respectively.
Our goal is to devise a provably consistent neural estimator of the DI rate from $\XX$ to $\YY$ based on a finite sample of these processes.
The section is organized as follows. We begin by demonstrating the difficulty of generalizing the MINE framework to the DI estimation. We then derive the DINE, discuss theoretical guarantees, and illustrate its implementation.

\subsection{Difficulties in Generalizing MINE to Directed Information}
Recall that the MINE \eqref{eq:mine_objective} is derived by approximating the potentials in the DV variational formula with NNs, and estimating expectations by sample means.
Generalizing to DI, we consider the conditional MI corresponding to the DI rate through $\lim_{n\to\infty}\sI(X^n;Y_n|Y^{n-1}) = \sI(\XX\to\YY)$ \cite{jiao2013universal}.
The corresponding KL term is given by
\begin{equation}
    \sI(X^n;Y_n|Y^{n-1}) = \DKL\big( \underbrace{P_{Y_n|Y^{n-1}X^n}}_{P_{\mathsf{pos}}}\|\underbrace{P_{Y_n|Y^{n-1}}}_{P_{\mathsf{neg}}} \big| P_{X^n Y^{n-1}} \big)\label{eq:di_mine_dkl},
\end{equation}
where $\DKL(P_{Y|X}\|Q_{Y|X}|P_{X})$ is the conditional KL divergence.
Estimating the expectations in DV representation of  \eqref{eq:di_mine_dkl} requires samples of both $P_{\mathsf{pos}}$ and $P_{\mathsf{neg}}$,
while only samples of $P_{X^nY^n}$ are available.
Samples of $P_{\mathsf{pos}}$ are the sampled channel outputs.
On the other hand, samples from $P_{\mathsf{neg}}$ require some manipulation of the data to break the relation between $\XX$ and $\YY$, but maintain temporal inter dependencies.
For i.i.d. data, random permutation of the samples is proposed \cite{belghazi2018mutual}, and for 1st order Markov processes the $1$-nearest neighbors algorithm is utilized  \cite{molavipour2021statistical}.
To the best of our knowledge, such a technique is unknown for unbounded memory,
as previous methods either affect both dependencies between $\XX$ and $\YY$ or the the temporal relations are restricted only to short memory.
As a solution, we derive a DV-based estimator of the DI rate that solely relies on samples from $P_{\mathsf{pos}}$, exploiting samples from an auxiliary distribution over $\cY$.

\subsection{The Estimator}\label{subsec:dine_derivation}
The DINE derivation relies on the following steps: First, we express DI as the difference between certain KL divergence terms. These are then represented via the DV variational formula (Theorem~\ref{theorem:DV}). Then, the DV potentials
are parametrized using RNNs, and expected values are approximated by sample means.
Recall that DI is given by
\begin{equation}
   \sI\left(X^n\to Y^n\right)=\sh\left(Y^n\right)-\sh\left(Y^n\|X^n\right).\label{eq:CC_ent_di_sec}
\end{equation}
For simplicity, assume $\cY$ is compact\footnote{This is a technical assumption that arises due to the choice of a uniform reference measure. By changing $P_{\tilde{Y}}$ to, e.g., Gaussian, this assumption is removed.}, and let $\tilde{Y}\sim\mathsf{Unif}(\cY)=:P_{\tilde Y}$ be independent of $\XX$ and $\YY$. Using the uniform reference measure we expand each entropy term as
\begin{subequations}
\begin{align}
    \sh(Y^n)&=\sh_{\mathsf{CE}}\left(P_{Y^n},P_{Y^{n-1}}\otimes P_{\widetilde{Y}}\right) - \DKL\left(P_{Y^n}\middle\| P_{Y^{n-1}}\otimes P_{\widetilde{Y}}\right)\label{eq:ent}\\
    \sh(Y^n \| X^n)&= \sh_{\mathsf{CE}} \left(P_{Y^n \| X^n},P_{Y^{n-1} \| X^{n-1}}\otimes P_{\widetilde{Y}}\middle|P_{X^n\|Y^{n-1}}\right) 
    \nonumber\\
    &\hspace{4cm}- \DKL\left(P_{Y^n\| X^n}\middle\| P_{Y^{n-1}\| X^{n-1}}\otimes P_{\widetilde{Y}}\middle|P_{X^n\|Y^{n-1}}\right),\label{eq:CC_ent}
\end{align}%
\end{subequations}
where $\sh_{\mathsf{CE}}(P_{Y|X},Q_{Y|X}|P_{X})$ is the conditional cross-entropy.
With some abuse of notation, let $\XX:=\{X_i\}_{i \in \mathbb{Z}}$ and $\YY:=\{Y_i\}_{i \in \mathbb{Z}}$ be the two-sided extension of the considered processes (the underlying stationary and ergodic measure remains unchanged). Inserting \eqref{eq:ent}-\eqref{eq:CC_ent} into \eqref{eq:CC_ent_di_sec} and using joint stationarity (which guarantees the existence of the following limit) we have
$$
    \sI(\XX \to \YY) =\sD_{Y\|X}^{\infty} - \sD_{Y}^{\infty} =\lim_{n \to \infty}\sD_{Y\|X}^{n} -  \lim_{n \to \infty}\sD_{Y}^{n},
$$
with
\begin{align*}\label{eq:dy_n_dxy_n}
    \sD_{Y\|X}^{n}&:=
    \DKL\left(P_{Y^0_{-(n-1)}\| X^0_{-(n-1)}}\middle\| P_{Y^{-1}_{-(n-1)}\| X^{-1}_{-(n-1)}}\otimes P_{\widetilde{Y}}\middle|P_{X^{0}_{-(n-1)}\|Y^{-1}_{-(n-1)}}\right)\\
    \sD_{Y}^{n}&:=\DKL\left(P_{Y^0_{-(n-1)}}\middle\| P_{Y^{-1}_{-(n-1)}}\otimes P_{\widetilde{Y}}\right).\numberthis
\end{align*}

To arrive at a variational form we make use of the DV theorem.
The optimal DV potentials for $\sD_{Y}^{n}$ and $\sD_{Y\|X}^{n}$ can be represented as dynamical systems that are given by the recursive relation
$
    z_{t+1} = f(z_t, u_t) 
$
for inputs $u_t$ and outputs $z_t$, respectively.
The dynamical system formulation follows from a representation of the optimal potentials in terms of the corresponding likelihood ratios.
As such, these potentials can be approximated to arbitrary precision by elements of the RNN function classes $\GRNN^{Y}$ and  $\GRNN^{XY}$ \cite{jin1995universal}.
The expectations in the DV formula are estimated with sample means (see Section \ref{subsec:consistency_proof}, where consistency of the DINE is proved, for details).
The DINE objective is given by
\begin{equation}
    \dine(\Dn,g_y,g_{xy}):=\hat{\sD}_{Y\|X}(\Dn,g_{xy})-\hat{\sD}_{Y}(\Dn,g_y)\label{eq:dine_objective},
\end{equation}
where
\begin{subequations}
\begin{align}
   \hat{\sD}_Y(\Dn, g_y) &:= \frac{1}{n}\sum_{i=1}^n{g_y}\left(Y^i\right)-\log\left(\frac{1}{n}\sum_{i=1}^n e^{g_y\left(\widetilde{Y}_i,Y^{i-1}\right)}\right),\label{eq:DINE_KL_est1}\\
   \hat{\sD}_{Y\|X}(\Dn,g_{xy}) &:= \frac{1}{n}\sum_{i=1}^n{g_{xy}}\left(Y^i,X^i\right)-\log\left(\frac{1}{n}\sum_{i=1}^n e^{g_{xy}\left(\widetilde{Y}_i,Y^{i-1},X^i\right)}\right).\label{eq:DINE_KL_est2}
\end{align}\label{eq:DINE_KL_est_unified}%
\end{subequations}
Consequently, the DINE is given by the optimization of \eqref{eq:dine_objective}
\begin{align}
    \dine(\Dn) &:=\sup_{g_{xy}\in\GRNN^{XY}}\hat{\sD}_{Y\|X}(\Dn,g_{xy})-\sup_{g_y\in\GRNN^{Y}}\hat{\sD}_{Y}(\Dn,g_y)\nonumber\\
    &=\sup_{g_{xy}\in\GRNN^{XY}}\inf_{g_y\in\GRNN^{Y}}\dine(\Dn,g_y,g_{xy}). \label{eq:dine_est_def}
\end{align}
The optimization can be executed via gradient-ascent over the RNN parameters.
 
\subsection{Theoretical Guarantees}\label{subsec:dine_theo}
The following theorem establishes consistency of the DINE.
\begin{theorem}[Theorem \ref{theorem:DINE_consistency}, restated]\label{theorem:DINE_consistency_in_guarantees}
Let $\XX$ and $\YY$ be jointly stationary and ergodic. Then for every $\epsilon>0$ there exists a positive integer $N$ such that for every $n>N$ we have
\begin{equation}
    \Big|\hspace{1mm}\dine(\Dn) - \sI(\XX\to\YY) \Big| \leq \epsilon, \qquad \PP-a.s.
\end{equation}
\end{theorem}
The proof can be divided into three main steps.
First, an \textit{information-theoretic} step, in which we express the DI rate as the difference of KL divergence terms, and represent it with the DV formula \eqref{eq:DV}.
Second, an \textit{estimation step}, that utilizes a generalization of Birkhoff's ergodic theorem \cite{breiman1957individual,algoet1988sandwich} to approximate the expectations of the DV representation by sample means.
Third, an \textit{approximation} step, in which we show that the sequence of optimal DV potential possesses a certain sequential structure, and approximate it using RNNs, utilizing a universal approximation theorem for RNNs \cite{jin1995universal}.
The proof is given in Section \ref{subsec:consistency_proof}.

\begin{remark}[Bound on the underlying DI rate]
The DINE is constructed as a difference of two maximization problems.
Therefore, while the DV representation induces a lower bound on each KL term for any choice of $g_y$ and $g_{xy}$, the overall objective \eqref{eq:dine_objective} does not bound the true DI-neither from above nor below. 
For a DINE variant that does bound $\sI(\XX\to\YY)$, one would need a variational upper bound of KL divergences that can be optimized over RNNs.
To the best of our knowledge, such a representation is not known.
\end{remark}

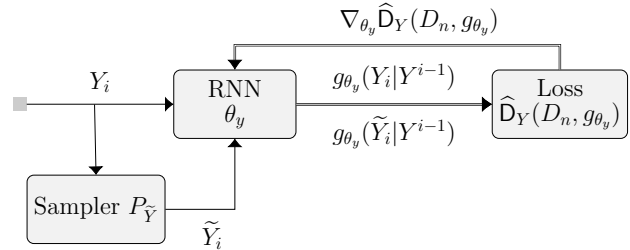
\begin{figure}[ht]
    \centering
     \scalebox{.65}{\input{Figures/dine_tikz}}
    \caption{The estimator architecture for the calculation of $\hat{\sD}_Y(\Dn, g_{\theta_{y}})$.}
    \label{fig:DINE_D_y}
\end{figure}

\subsection{Implementation}\label{subsec:dine_implement}
We describe the implementation details of the DINE.
Fix $k_{y}$ and $k_{xy}$ with the corresponding RNN classes $\GRNN^{(d_y,1,k_y)}$ and $\GRNN^{(d_y,1,k_{xy})}$.
The corresponding compact parameter subsets are denoted $\Theta_{y}\subseteq\RR^{d_{\theta_y}}$ and $\Theta_{xy}\subseteq\RR^{d_{\theta_{xy}}}$ with finite $d_{\theta_y}$ and $d_{\theta_{xy}}$.
The RNNs over which we optimize comprise a \emph{modified} long short-term memory (LSTM) layer and a fully connected (FC) network.
We consider 
The architecture for $\hat{\sD}_{Y}(\Dn, g_{\theta_{y}})$ is depicted in Figure \ref{fig:DINE_D_y}.
We next present the modified LSTM cell, discuss the optimization procedure, and propose an adjustment for the DINE objective that accounts for possible estimation variance induced by the reference samples.

\subsubsection{\texorpdfstring{\underline{Modified LSTM}}{Modified LSTM}}
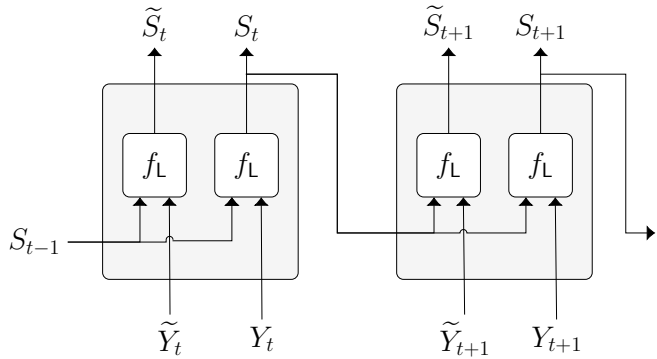
\begin{figure}[ht]
    \centering
     \scalebox{.65}{\input{Figures/modified_lstm}}
    \caption{The modified LSTM cell unrolled in the DINE architecture of $\widehat{\sD}_{Y}$. Recursively, at each time $t$, $(Y_t, S_{t-1})$ and $(\widetilde{Y}_t, S_{t-1})$ are mapped to $S_t$ and $\widetilde{S}_t$, respectively.}
    \label{fig:modified_lstm}
\end{figure}

Note that the RNN mappings in each KL estimate in \eqref{eq:DINE_KL_est_unified} consists of the same mapping, each time differing on the $i$th input.
Our goal is therefore to construct a unified mapping for samples of both the joint and reference distribution, while restricting memory to depend only on past samples from the joint distribution.
To this end, we adjust the structure of the classic LSTM cell \cite{hochreiter1997long}.
The modification is presented for $\hat{\sD}_{Y}$ and is straightforwardly adopted for $\hat{\sD}_{Y\|X}$.
The classic LSTM is an RNN that recursively computes a hidden state $s_i$ from its input $y_i$ and the previous state $s_{i-1}$, (see \cite{hochreiter1997long} for more background on LSTM).
We henceforth use the shorthand $s_i=f_{\mathsf{L}}(y_i,s_{i-1})$ for the relation between $s_i$ and $(y_i,s_{i-1})$ defined by the LSTM.
As DINE also employs the sequence $\tilde{y}^n$ drawn from the reference distribution $P_{\tilde{Y}}$, the modified LSTM collects hidden states for both $y^n$ and $\tilde{y}^n$.
At time $i=1,\ldots,n$, the cell takes a pair $(y_i,\tilde{y}_i)$ as input, and outputs two hidden states $s_i= f_{\mathsf{L}}(y_i, s_{i-1})$ and $\widetilde{s}_i = f_{\mathsf{L}}(\widetilde{y}_i, s_{i-1})$, with only $s_i$ passed on for calculating the next state.
The state sequences are then processed by the FC network to obtain the elements of \eqref{eq:DINE_KL_est1}.
The states $s_i$ and $\tilde{s}_i$ calculate a summary of $y^{i-1}$ and $\tilde{y}^{i-1}$ through the LSTM cell recursive mapping.
Therefore, we interpret the computation of $g_{\theta_y}(y^i)$ and $g_{\theta_y}(\tilde{y},y^{i-1})$ as conditioning on past inputs.
With some abuse of notation, we use interchangeably the following conditional form for the DINE outputs.
\begin{align*}
    g_{\theta_y}(y^i) &=  g_{\theta_y}(y_i,s_{i-1})= g_{\theta_y}(y_i|y^{i-1})\\
    g_{\theta_y}(\tilde{y}_i,y^{i-1}) &= g_{\theta_y}(\tilde{y}_i,s_{i-1}) = g_{\theta_y}(\tilde{y}_i|y^{i-1})\numberthis{}{}\label{eq:dine_cond_form}.
\end{align*}
The notation on the right-hand sides (RHSs) of \eqref{eq:dine_cond_form} emphasizes that the input dimension is fixed for each time step.
The calculation of the hidden states for $\hat{\sD}_{Y\|X}$ in \eqref{eq:DINE_KL_est2} is performed analogously, by replacing $y_i$ and $\tilde{y}_i$ with $(x_i,y_i)$ and $(x_i,\tilde{y}_i)$, respectively.
The modified LSTM cell is shown in Figure \ref{fig:modified_lstm}.

\subsubsection{\texorpdfstring{\underline{Algorithm}}{Algorithm}}
The DINE algorithm computes $\dine(\Dn)$ by optimizing the parameters $\theta_{y}\in\Theta_y$ and $\theta_{xy}\in\Theta_{xy}$ of the RNNs $g_{\theta_y}$ and $g_{\theta_{xy}}$, respectively. 
We divide the dataset into batches of $B$ sequences of length $T$, i.e., $\Bm:= (X^m,Y^m)$ with $m=BT<n$.
For each batch, we provide samples of the reference measure\footnote{In practice, we sample uniformly from the smallest $d$-dimensional bounding hypercube of the samples $Y^n$.} and feed the sequences through the DINE architecture to obtain the DV potentials $g_{\theta_y}$ and $g_{\theta_{xy}}$.
Those are then used to calculate the DINE objective \eqref{eq:dine_objective}, from which gradients are derived for the update of $\theta_y$ and $\theta_{xy}$.
We repeat the above steps until some convergence criteria is met.
See Algorithm \ref{alg:DINE} in Section \ref{sec:main_results} for the full list of steps.
The weights of the FC networks within each RNN are shared since we wish to produce the same function acting on different inputs.

\subsubsection{\texorpdfstring{\underline{Reference samples}}{Reference samples}}
The exponential terms in \eqref{eq:DINE_KL_est1} and \eqref{eq:DINE_KL_est2} can potentially cause instability in the estimation process by biasing the estimate of the update gradients \cite{belghazi2018mutual}.
Existing methods to account for this problem include moving average filtering of the gradients \cite{belghazi2018mutual} and clipping of the exponential terms \cite{song2019understanding}.
Herein, we exploit the reference uniform measure.
For each $i$, we collect $K_U$ reference samples $\{\tilde{Y}_{i,j}\}_{j=1}^{K_U}$.
These are used to calculate the corresponding DV potentials by averaging over the reference samples,
$$
    \bar{g}_{\theta_y}(\tilde{Y}_i|Y^{i-1}) := \frac{1}{K_U}\sum_{j=1}^{K_U} e^{g_y(\tilde{Y}_{i,j}|Y^{i-1})},\quad
    \bar{g}_{\theta_{xy}}(\tilde{Y}_i|Y^{i-1},X^i) := \frac{1}{K_U}\sum_{j=1}^{K_U} e^{g_{xy}(\tilde{Y}_{i,j}|Y^{i-1},X^i)}.
$$
We then use $\bar{g}_{\theta_y}$ and $\bar{g}_{\theta_{xy}}$ instead of the aforementioned exponential terms in \eqref{eq:DINE_KL_est1} and \eqref{eq:DINE_KL_est2}.
We observe empirically that the averaging reduces bias and numerical instability in the estimation process.

\section{DINE Optimization Over Continuous Spaces}\label{sec:cont_opt}
In this section we present our method for the optimization of the DINE over continuous input distributions.
We utilize a generative model, whose objective is to construct a sample $\Dn$ that maximizes \eqref{eq:dine_est_def}.
In what follows, we derive the optimizer, discuss its theoretical properties, describe its implementation, and discuss the joint estimation-optimization procedure.

\subsection{Optimizer Derivation}
We consider the optimization
$
    \sup_{\sP_{X}}\sI(\XX\to\YY),
$
where $\sP_{X} = \{P_{X_i|X^{i-1}}\}_{i\in\NN}$ for feedforward channels and $\sP_{X} = \{P_{X_i|X^{i-1}Y^{i-1}}\}_{i\in\NN}$ for channels with feedback.
To that end, we propose the NDT, an RNN-based generative model that maps an arbitrary i.i.d. sequence, $U^n\sim P^{\otimes n}_U$, to a sequence of channel inputs.
The NDT is given by $\NDT\in\GRNN^X=\GRNN^{(d_x,d_x,k)}$ with parameters $\phi\in\Phi$.
Recall that $\NDT$ recursively calculates the sequence of channel inputs $X^{\phi,n}$, where $X^\phi_0=0$ and
\begin{equation}\label{eq:ndt_samples_gen}
   X^\phi_i=\NDT(U_i,Z^\phi_{i-1}),\quad i=1\dots,n.
\end{equation}
The sequence $X^{\phi,n}$ is passed through the channel to obtain the corresponding outputs $Y^{\phi,n}$, to arrive at the dataset $\Dphi(U^n):=(X^{\phi,n},Y^{\phi,n})$.
For feedforward channels we take $Z^\phi_i = X^\phi_i$, while $Z^\phi_i = (X^\phi_i,Y^\phi_i)$ for channels with feedback. 
To simplify notation, we denote $\Dphi(U^n) = \Dphi$ and consider the same distribution $P_U$ throughout.
The overall optimization is given by
\begin{equation}
    \dine^\star(U^n):=\sup_{\NDT\in\GRNN^X}\dine(\Dphi, \NDT) = \sup_{\NDT\in\GRNN^X}\left(\sup_{g_{xy}\in\GRNN^{XY}}\inf_{g_y\in\GRNN^{Y}}\dine(\Dphi, \NDT, g_y, g_{xy})\right)\label{eq:ndt_opt}.
\end{equation}
The DINE objective \eqref{eq:dine_objective} acts as a loss function for the optimization of $\NDT$, which is executed via gradient-based optimization over $\phi$.
When the channel is memoryless we focus on MI estimation and optimization, employing the MINE.
Consequently, $\NDT$ only takes $U_i$ as input and the optimization is carried out over $\cG_k^{(d_{x},d_{x})}$. 
We next inspect the theoretical properties of the combined estimation-optimization method.

\subsection{Theoretical Guarantees}\label{subsec:ndt_theo}
In this section we provide theoretical analysis of the performance and structure of the proposed method.
We first account for the convergence of the joint optimization procedure.
Then, restricting attention to MI optimization for memoryless channels, we characterize the optimized NDT structure.

\subsubsection{\texorpdfstring{\underline{Consistency}}{Consistency}}\label{subsec:dine_ndt_consistency}
We show that under appropriate assumptions on the channel transition kernel and input distribution, the optimization in \eqref{eq:ndt_opt} converges to the maximal DI.
We begin by describing the class of channel inputs that our result accounts for.
We consider the class of stationary processes $\XX$, for which there exist an auxiliary stationary process $\BS$ over $\cS\subseteq\RR^{d'}$ and a function $f_\mathsf{s}\in\cC(\cX\times\cS)$ such that 
$$
S_i = f_\mathsf{s}(H_i, S_{i-1}),\quad i\in\NN,
$$
and $H^{i-1} \leftrightarrow S_{i-1}  \leftrightarrow X_i$ forms a Markov chain.
We take $H_i = X_i$ for computing the feedforward capacity and $H_i = (X_i,Y_i)$ for the feedback capacity.
We call such processes \emph{recursive-state processes} (RSPs) and denote the class of RSPs by $\sX_{\cS}$.
In Section \ref{sec:ndt_consistency_proof}, where the consistency of the DINE-NDT method is proved, we show that $\sX_{\cS}$ can be represented as a special case of the general state-space model \cite[Eqn.~(3.1)-(3.2)]{moller2011stochastic} by constructing a functional reformulation of the aforementioned Markov relation.
The structure allows $f_{\mathsf{s}}$ to be a randomized function.
To better understand the breadth of the class $\sX_{\cS}$, we make the following observation.
\begin{lemma}\label{lemma:markov_subset_rsp}
The class of stationary Markov processes of finite order is a subset of $\sX_{\cS}$.
\end{lemma}
The proof is straightforward by choosing $ S_i = [X_{i-(m-1)},\dots,X_i]$, for $ i\geq m-1$, with Markov order $m$. When $i< m-1$, the $i$th to $(m-1)$th entries are zeros.

We next describe the considered class of channels.
A unifilar state channel (USC) \cite[Section~2]{ziv1985universal} is a channel whose latent state $Z_i$ evolves according to
$$
  Z_i = f_\mathsf{z}(Z_{i-1},Y_i,X_i),\quad i\in\NN,
$$
for some $f_\mathsf{z}\in\cC^1(\cZ\times\cX\times\cY)$, where $(X^{i-1},Y^{i-1}) \leftrightarrow(X_i,Z_{i-1},Y_i) \leftrightarrow Z_i$ forms a Markov chain.
We consider USCs with continuous input and output spaces, whose outputs adhere to the functional relation
$$
    Y_i = f_\mathsf{y}(Z_i,X_i,K_i),
$$
for some $f_\mathsf{y}\in\cC^1(\cZ\times\cX\times\cY)$ and an i.i.d. external process  with $K_1\sim P_{K}\in\lebmeas(\RR^{d_K})$ for some $d_K\in\NN$.
This structure can be viewed as a variation of \cite[Equation 7]{kim2008coding}, in which the channel mapping also receives past outputs and the state is unifilar.
To bound the effective estimation-optimization error, we impose the following Lipschitz condition on the functions $f_\mathsf{z},f_\mathsf{y}$.
\begin{assumptions}\label{assumption:dine_ndt_channel_assumption}
$f_\mathsf{z}$ and $f_\mathsf{y}$ are Lipschitz continuous with Lipschitz constants $M_y$ and $M_z$, respectively, such that $M_y(M_z+1)<1$.
\end{assumptions}
This assumption can be lifted if we do not permit any recursive relation in the channel structure (for more details, see Sec. \ref{sec:ndt_consistency_proof}).
In addition, we assume that the DINE RNNs, $(g_y, g_{xy})$, are Lipschitz with some finite Lipschitz constants $M_1$ and $M_2$.
We have the following consistency claim.
\begin{theorem}[Theorem \ref{theorem:ndt_consistency0}, restated]\label{theorem:ndt_consistency}
Fix $\epsilon>0$, let $U^n\sim P^{\otimes n}_U$ and consider the continuous USC $\{P_{Y_i|Y^{i-1},X^i}\}_{i\in\NN}$, where $f_\mathsf{y}, f_\mathsf{z}$ satisfy Assumption \ref{assumption:dine_ndt_channel_assumption}. Let $\underline{\sC}_s$ be the supremum of the DI rate $\sI(\XX\to\YY)$ over $\sX_{\cS}$. Then, there exists $N\in\NN$ such that for every $n>N$, we have
\begin{equation}
    \left|\underline{\sC}_s-\dine^\star(U^n)\right|\leq\epsilon,\qquad \PP-a.s.,
\end{equation}
where $\dine^\star(U^n)$ is given in \eqref{eq:ndt_opt}.
\end{theorem}

The proof is given in Section \ref{sec:ndt_consistency_proof}, where we also generalize this statement to the feedback scenario.
The proof utilizes tools such as FRL and universal approximation for RNNs.
\begin{remark}[Feasible channels]
In general, $\underline{\sC}_s$ lower bounds the capacity of a given channel with memory, and the characterization of capacity-achieving input distributions of arbitrary stationary channels with continuous input and output spaces is currently an open problem.
However, when the channel is Gaussian and the channel has a linear state-space model, the capacity achieving distribution can be reformulated as an RSP \cite{sabag2021feedback, pedram2018some}.
\end{remark}

\subsubsection{\texorpdfstring{\underline{Optimized NDT structure}}{Optimized NDT structure}}\label{subsec:opt_ndt}
We now restrict attention to memoryless channels and thus focus on MI estimation and optimization.
We employ MINE as the MI estimator and discuss the structure of the optimized NDT.
To this end, we utilize a multivariate generalization of the CDF, originally proposed by Kn\"othe \cite{knothe1957contributions} and Rosenblatt \cite{rosenblatt1952remarks}.
Consider a $d$-dimensional random vector $X:=(X_1,\dots,X_d)\sim P_X\in\cP(\cX)$, where $\cX\subseteq\RR^d$ and define the associated vector-valued function $T_X:\cX\to [0,1]^d$ by
\begin{equation}
    \begin{split}
    \big[T_X(x)\big]_1&= \PP(X_1\leq x_1)\\
    \big[T_X(x)\big]_i&=\PP\Big(X_i\leq x_i\,\Big|\big[T_X(X)\big]_{i-1}=\big[T_X(x)\big]_{i-1},\ldots, \big[T_X(X)\big]_1=\big[T_X(x)\big]_1\Big),\quad i=2,\ldots d.    
    \end{split}\label{eq:d_cdf_map}
\end{equation}
In words, for $x\in\cX$, each entry $\big[T_X(x)\big]_i$ is given by the conditional distribution function of $X_i$ at $x_i$ given the values of the function in the preceding entries, i.e., $\big[T_X(x)\big]_1,\ldots,\big[T_X(x)\big]_{i-1}$. We have the following proposition.
\begin{lemma}\label{lemma:gen_inverse_transform}
Let $X\sim P_X\in\lebmeas(\cX)$ with $\cX\subseteq\RR^{d_x}$ and consider the map $T_X:\cX\to[0,1]^{d_x}$ defined above. Then,
\begin{enumerate}
    \item The function $T_X$ is Borel measurable and the random variable $T_X(X)$ is uniformly distributed over $[0,1]^{d_x}$.
    \item $T_X$ is a bijection and for $U\sim\mathsf{Unif}\big([0,1]^{d_x}\big)$, we have $T_X^{-1}(U)\stackrel{d}{=}X$.
\end{enumerate}
\end{lemma}
The proof of Lemma \ref{lemma:gen_inverse_transform} is in Appendix \ref{proof:gen_inverse_transform_proof}. 
The measurability of $T$ follows from its definition as a vector-valued right continuous function, the distribution of $T_X(X)$ follows from the definition of $T$, bijectivity follows from the positive semi-definite property of the Jacobian of $T$. The distribution of $T_X^{-1}(U)$ follows from the tools developed in \cite{huang2018neural} in the context of neural autoregressive flows. 
The reader is referred to \cite{hyvarinen1999nonlinear,irons2021triangular,spantini2018inference} for  further discussion and useful properties of the function $T_X$.

\medskip
Lemma \ref{lemma:gen_inverse_transform} provides a representation of continuous random variables as functions of uniformly distributed variables\footnote{As a consequence of Lemma \ref{lemma:gen_inverse_transform}, we can construct a transformation between any two absolutely continuous random variables $W$ and $X$ provided they have the same dimension, by utilizing the composition $T_X^{-1}\circ T_W:\cW\mapsto\cX$ \cite{papamakarios2021normalizing}.}. We leverage this fact to characterize to MINE-maximizing NDT. 
Let $\cP_p(\RR^d)$ be the class of Borel probability measures on $\RR^d$ with finite $p$-th moment, i.e., $\int \|x\|^p \, \dd \mu(x)<\infty$.
For a given transition kernel $P_{Y|X}$, let $\sC$ denote the capacity of the corresponding memoryless channel bound to a second moment input constraint.
Denote the capacity-achieving distribution $P_{X^\star}:=\argmax_{P_X\in\cP_2(\cX)}\sI(X;Y)$, let  $X^\star\sim P_{X^\star}$, and consider its associated mapping $T_{X^\star}$. 
We quantify the distance between the NDT-induced probability distribution and $P_{X^\star}$ using the $2$-Wasserstein distance. 
The $p$-Wasserstein distance between $\mu,\nu\in\cP_p(\RR^d)$ is given by 
$$
\sW_p(\mu,\nu):=\left[ \inf_{\pi\in\Pi(\mu,\nu)}\int_{\RR^d\times\RR^d}\|x-y\|^p\dd \pi(x,y)\right]^{1/p},
$$
where $\Pi(\mu,\nu)$ is the set of all couplings of $\mu$ and $\nu$. We propose the following theorem.
\begin{theorem}[Optimal NDT]\label{theorem:opt_ndt}
Fix $\epsilon>0$ and let $P_U$ be the uniform distribution over $\cU$. 
Let $P_{Y|X}$ with a bounded and continuous PDF $p_{Y|X}$ such that it induces a finite second moment on the channel output for any second moment-bounded input distribution. 
Then, there exist $\NDT\in\GNN^{(d_x,d_x)}$, such that 
\begin{equation}
    \sW_2\left(\NDTpush P_U,P_{X^\star}\right)\leq \epsilon,\label{eq:ndt_struct_dkl_bound}
\end{equation}
where $\NDTpush P_U$ is the pushforward measure of $P_U$ by $\NDT$.
Moreover, there exist $n_0\in\NN$ such that for any $n>n_0$ and $U^n\sim P^{\otimes n}_U$, we have
\begin{equation}
    \left|\hspace{1mm}\sC - \mine\big(\Dphik\big) \right|\leq \epsilon,\qquad \PP-a.s.\label{eq:ndt_struct_mine_bound},
\end{equation}
where, $\Dphi:=\{(h_{\phi}(U_i), Y_i)\}_{i=1}^n$.
\end{theorem}
The proof of Theorem \ref{theorem:opt_ndt} is in Section \ref{subsec:proof_ndt_strcut}.
It follows by approximating both distributions with a smoothed version of them, obtained by a convolution with a sequence of isotropic Gaussian distributions with decreasing variance.
We then argue for the convergence of the $p$-Wasserstein metric and capacity estimate by the weak continuity of the $p$-Wasserstein distance, Wasserstein continuity of KL-divergence \cite{polyanskiy2016wasserstein}, weak continuity of differential entropies for distributions with bounded moments \cite[Theorem~1]{godavarti2004convergence} and the MINE consistency \cite[Theorem~2]{belghazi2018mutual}.
Theorem \ref{theorem:opt_ndt} guarantees the existence of an NDT model that approximates the capacity achieving distribution (under the $p$-Wasserstein distance), which, in turn, yields a consistent MINE-based proxy of capacity. 
We therefore conjecture
that the MINE-maximizing NDT is in fact an approximator of $T_X^{-1}$.
We empirically validate this conjecture for the AWGN channel in the next section.

\begin{remark}[Lower bounding channel capacity] \label{remark:capacity_lb_mine}
When the MINE is optimized, the NDT does not impede the DV-induced lower bound (Lemma \ref{lemma:mine_lb}).
Consequently, for any $\NDT\in\GNN^{(d_x,d_x)}$, the corresponding MINE output lower bounds the channel capacity.
This property will serve us in Section \ref{subsec:amp_const_awgn} to propose a bound on the capacity of the peak-power constrained AWGN.
\end{remark}

\begin{figure}[t]
    \centering
     \scalebox{.75}{\input{Figures/ndt_tikz}}
    \caption{The NDT. The noise and past channel output (if feedback is present) are fed into an RNN. The last layer imposes a constraint of our choice.}
    \label{fig:NDT_architecture}
\end{figure}
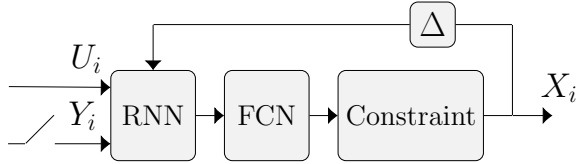

\subsection{Implementation}
\begin{figure}[ht]
    \centering
     \scalebox{.62}{\input{Figures/DINE_NDT}}
    \caption{The complete system for optimization over continuous spaces. On each step gradients are passed to a predetermined model, while the other one's parameters are fixed.}
    \label{fig:complete_system}
\end{figure}
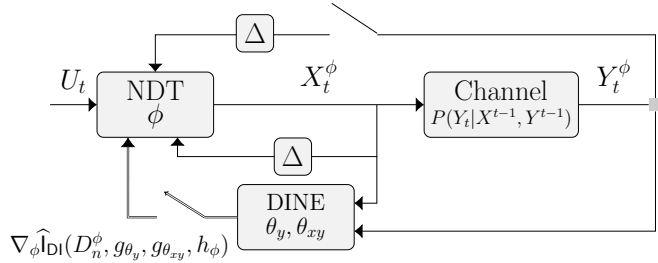

The NDT is implemented using an LSTM stacked with 2 FC layers.
Chanel input constraints, such as average or peak-power constraints, can be imposed on the NDT outputs, as long as these can be realized with a differential function of the $\phi$.
The NDT model is shown in Figure \ref{fig:NDT_architecture}.
The overall optimization over the NDT and DINE takes the form
$$
\sup_{\phi\in\Phi,\theta_{xy}\in\Theta_{xy}}\inf_{\theta_{y}\in\Theta_{y}}\dine(\Dphi,g_{\theta_{y}},g_{\theta_{xy}},\NDT).
$$
In every iteration, we draw a noise batch $(U^m)$, from which $\Bphi = (X^{\phi,m},Y^{\phi,m})B$ is computed.
The batch $\Bm$ is processed by $g_{\theta_y}$ and $g_{\theta_{xy}}$, the loss $\widehat{\sI}(\Bphi,g_{\theta_y},g_{\theta_{xy}},\NDT)$ is calculated, and gradients are propagated to update the models weights. 
Figure \ref{fig:complete_system} illustrates the complete architecture.

The training adheres to an alternating optimization procedure.
Namely, we iterate between updating $(\theta_y, \theta_{xy})$ and $\phi$, each time keeping the other parameters fixed.
After the training is done, we perform a long Monte-Carlo (MC) evaluation to obtain an estimate of \eqref{eq:dine_objective}.
The procedure is summarized in Algorithm \ref{alg:cap_est} and its implementation is available on \href{https://github.com/DorTsur/dine_ndt}{GitHub}.
This alternation between two models sharing a common loss is found in other fields, such as generative adversarial networks \cite{goodfellow2014generative} and actor-critic algorithms \cite{konda2000actor}.
We stress that the proposed optimization scheme can be applied to any NN-based estimator of information measures, inasmuch as it is differentiable w.r.t. the NDT outputs.

\section{Empirical Capacity Estimation Results}\label{sec:cont_opt_exp}
We demonstrate the performance of Algorithm \ref{alg:cap_est} for continuous channel capacity estimation, considering both feedforward and feedback scenarios for several channel models. The numerical results are compared with the available theoretical solution/bounds to verify the effectiveness of the proposed method.
The simulations are implemented in TensorFlow \cite{tensorflow2015}.
The DINE is implemented using a modified LSTM and two fully-connected layers with 50, 100 and 50 neurons, respectively.
The NDT is implemented with an LSTM and two fully connected layers, each with 100 neurons, stacked with an output layer with $d_x$ neurons. 

We note that the term calculated by the DINE-NDT method differs from the general capacity expression in the following way.
In \eqref{eq:ndt_opt}, we take the supremum over the estimated DI \textit{rate}, i.e., the limit is taken before the supremum.
In contrast, the general capacity expression \eqref{eq:CFF} considers the opposite order of limit and optimization. This order is known to be interchangeable for stationary Gaussian channels \cite{kim2009feedback}, and generally seems to have a minimal effect on the accuracy of the numerical results for the considered examples.
We also stress that all methods with which we compare the DINE-NDT method assume full knowledge of the channel model, which our approach does not require.

\subsection{AWGN Channel}
\subsubsection{\texorpdfstring{\underline{Average power constraint}}{Power Constraint}}
We consider the AWGN channel
\begin{equation}
    Y_i = X_i + Z_i,  \quad i \in \mathbb{N},
\end{equation}
where $Z_i \sim \mathcal{N}\left(0, \sigma^2\right)$ are i.i.d. and $ X_i $ is the channel input sequence bound to the average power constraint $\mathbb
E\left[ X_i^2 \right] \leq P$. The capacity of this channel is given by $\mathsf{C} = \frac{1}{2}\log\left(1+\frac{P}{\sigma^2}\right)$~\cite{CovThom06}. 
We set $\sigma^2 = 1$ and estimate the capacity via the optimized MINE for a range of $P$ values. The numerical results are compared to the analytic solution in Figure \ref{fig:awgn_results}, where a clear correspondence is seen.

\subsubsection{\texorpdfstring{\underline{Peak power constraint}}{Peak-Power constraint}}\label{subsec:amp_const_awgn}
\begin{figure}[t]
  \begin{subfigure}[t]{.5\textwidth}
    \centering
    \includegraphics[trim={31pt 1pt 3pt 20pt}, clip, width=0.8\linewidth]{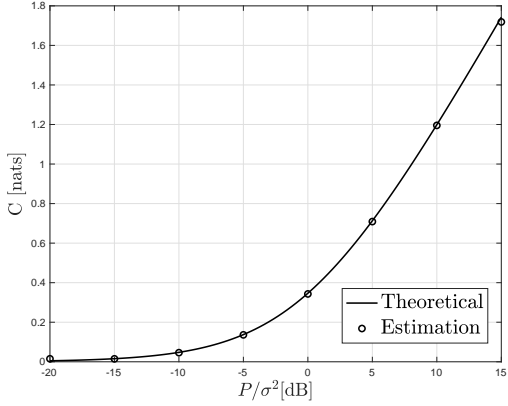}
    \caption{Average power constraint.}
    \label{fig:awgn_results}
  \end{subfigure}
  \hfill
  \begin{subfigure}[t]{.5\textwidth}
    \centering
    \includegraphics[trim={27pt 1pt 20pt 26.4pt}, clip, width=0.78\linewidth]{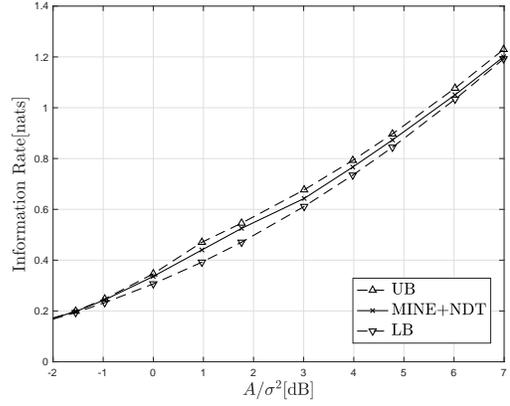}
    \caption{Peak power constraint.}
    \label{fig:awgn_amplitude}
  \end{subfigure}

  \medskip

  \begin{subfigure}[t]{.5\textwidth}
    \centering
    \includegraphics[trim={20pt 1pt 4pt 26.4pt}, clip, width=0.8\linewidth]{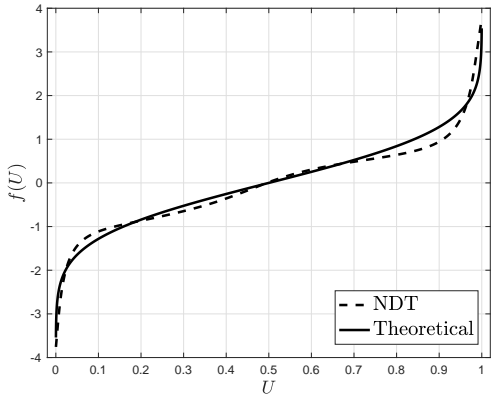}
    \caption{Optimized NDT structure comparison with $F_X^{-1}$.}
    \label{fig:ndt_struct}
  \end{subfigure}
  \hfill
  \begin{subfigure}[t]{.5\textwidth}
    \centering
    \includegraphics[trim={40pt 21pt 20pt 26.4pt}, clip, width=0.8\linewidth]{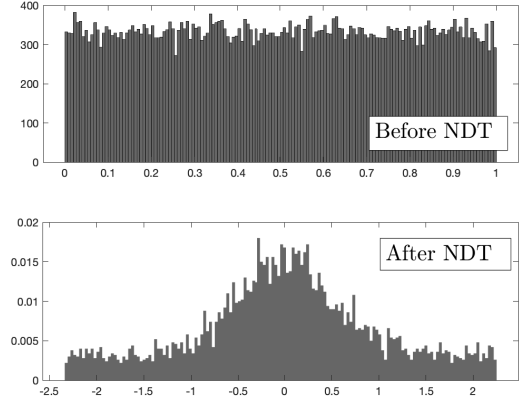}
    \caption{NDT input vs. output histogram.}
    \label{fig:ndt_hist}
  \end{subfigure}
  \caption{Performance of the proposed method in the AWGN channel for both (a) average and (b) peak power constraints.
  For the average power constrained AWGN with $P=1$, we compare the optimized NDT structure with $T_{X^\star}^{-1}$ (c), and present the NDT output for a set of independent uniform samples (d).}
\end{figure}

We consider the AWGN channel with a peak power constraint $|X|<A$, for some $A>0$.
The capacity of this channel is unknown, but upper and lower bounds on it are available in the literature \cite{ozarow1990capacity,thangaraj2017capacity}. 
In Figure \ref{fig:awgn_amplitude} we present a comparison of the capacity estimate obtained from our Algorithm \ref{alg:cap_est} (with MINE instead of DINE) and the aforementioned bounds. Evidently, the estimate falls within the theoretical bounds. As MINE lower bounds the channel capacity for any choice of $\NDT$ (cf., Remark \ref{remark:capacity_lb_mine}), our estimate also provides new and tighter lower bounds on the capacity of this channel.

\subsubsection{\texorpdfstring{\underline{Optimized NDT structure}}{Optimized NDT structure}}
Considering the average power constrained AWGN, we check two characteristics of the MINE-maximizing NDT.
First, we empirically validate Theorem \ref{theorem:opt_ndt} by comparing the optimized NDT with $T^{-1}_{X^\star}$, where $X^\star\sim\cN(0,P)$ is the capacity-achieving input.
The correspondence is shown in Figure \ref{fig:ndt_struct}. Second, in Figure \ref{fig:ndt_hist} we examine histograms to further verify that the optimized NDT maps the input samples $U^n$ into samples of the capacity-achieving Gaussian distribution.

\subsection{Gaussian MA(1) Channel}
\begin{figure}[t]
    \centering
    \begin{subfigure}[b]{0.32\textwidth}
        \centering
        \includegraphics[trim={55pt 10pt 20pt 26.4pt}, clip,width=1\textwidth]{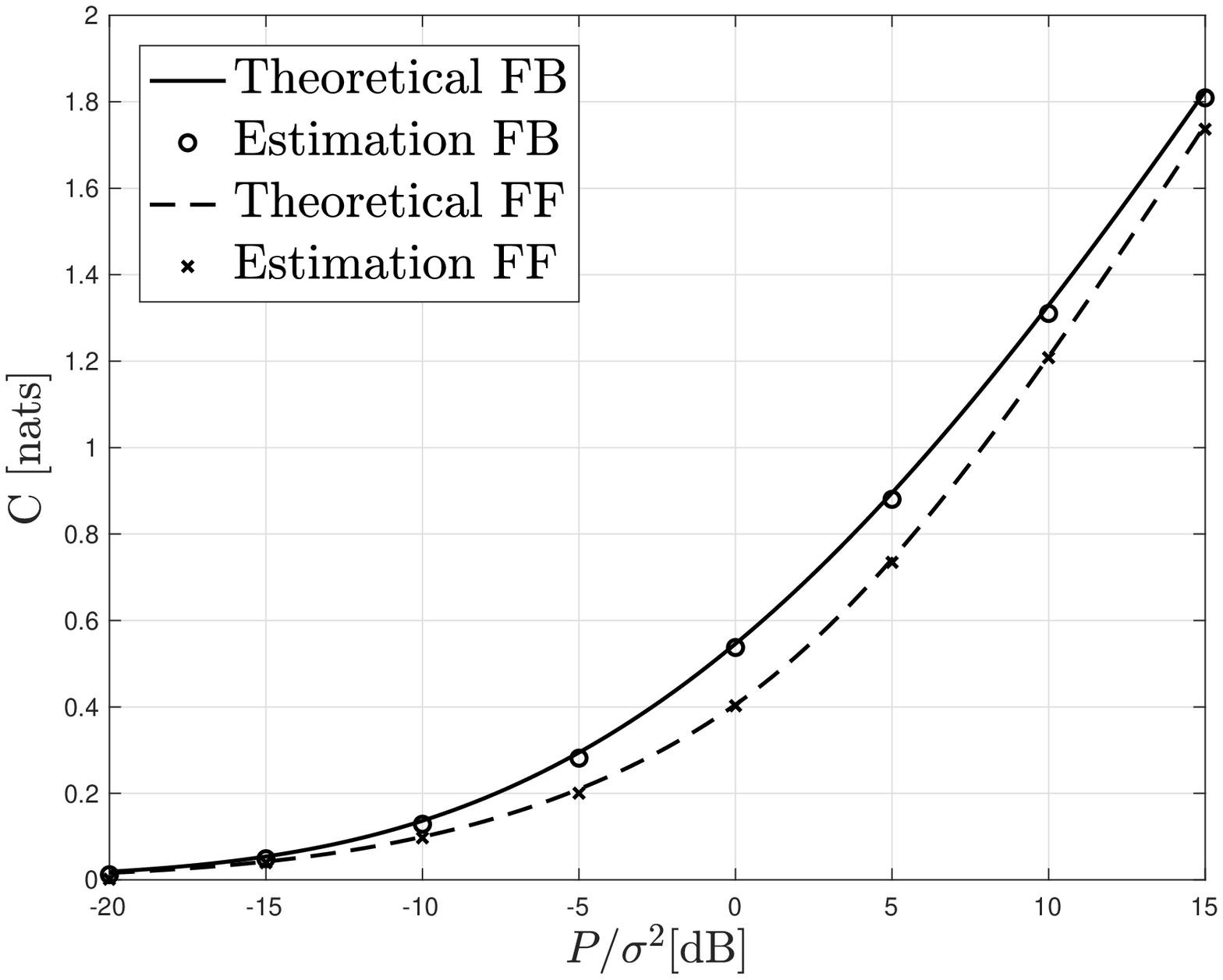}
        \caption{MA(1)-AGN capacity.}
        \label{fig:C_magn_comp}
    \end{subfigure}
    \begin{subfigure}[b]{0.32\textwidth}
        \centering
        \includegraphics[trim={55pt 10pt 20pt 26.4pt}, clip, width=\linewidth]{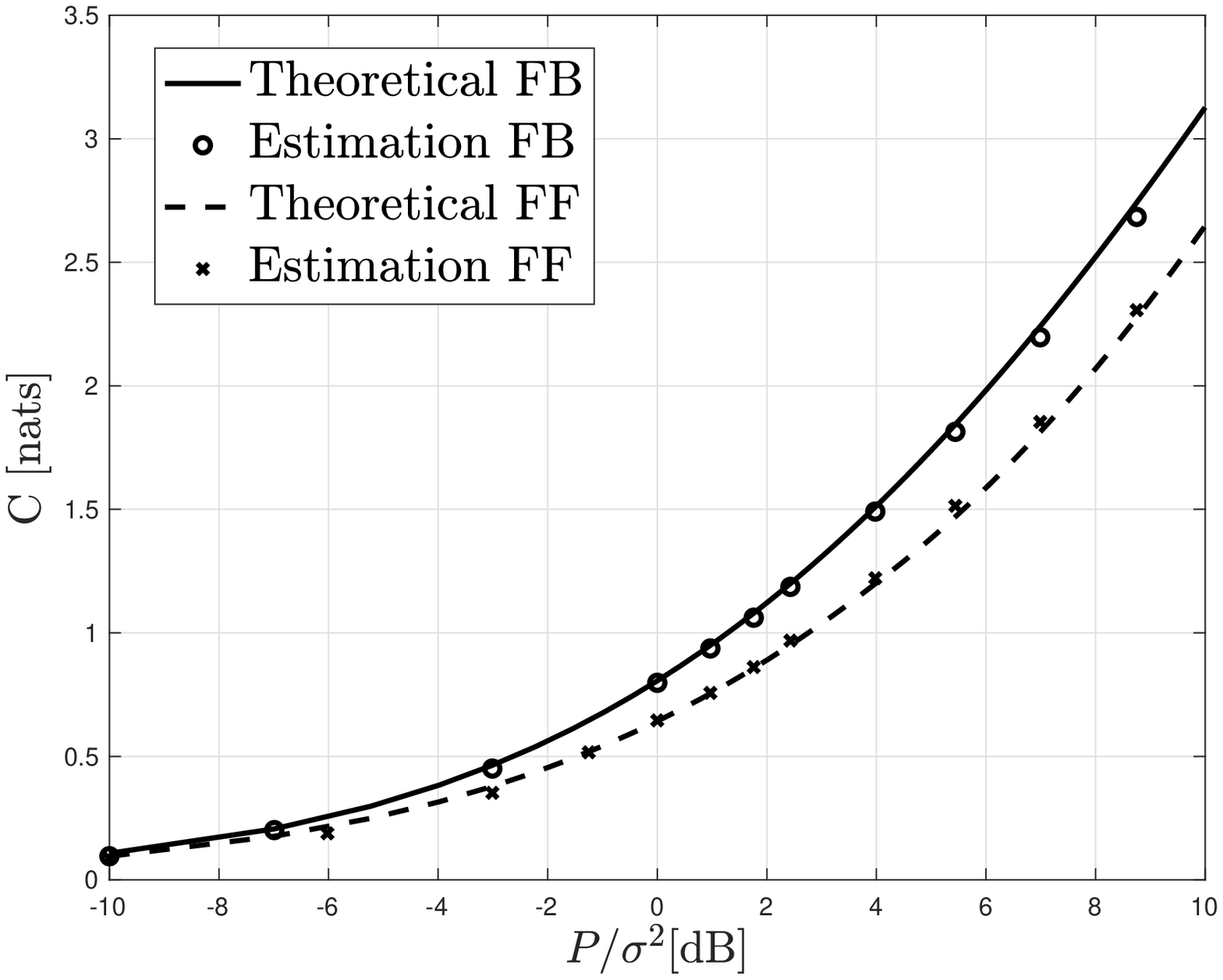}
        \caption{MIMO AR(1)-AGN capacity.}
    \label{fig:ar_4d}
    \end{subfigure}
    \begin{subfigure}[b]{0.32\textwidth}
        \centering
        \includegraphics[trim={55pt 10pt 20pt 26.4pt}, clip,width=1\textwidth]{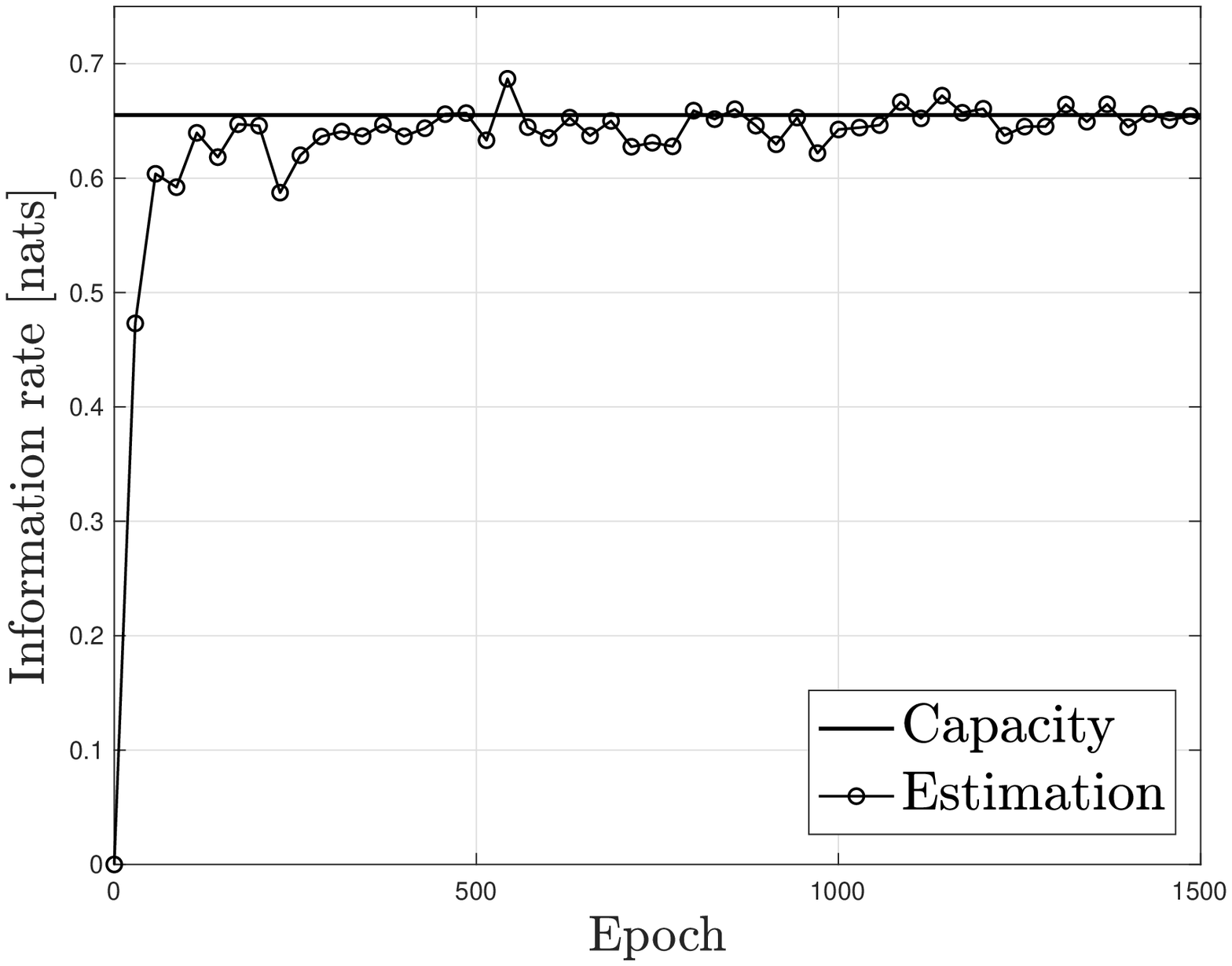}
        \caption{Algorithm convergence}
        \label{fig:ar_4d_convergence}
    \end{subfigure}
    \caption{Capacity estimation results for Gaussian channels with memory.
     Figure (a) presents capacity estimation results for the MA(1)-AGN channel.
     Figures (b) and (c) refer to the MIMO AR(1)-AGN channel, presenting both feedforward and feedback capacities for a variety of SNR values, and demonstrating the algorithm convergence for $P=1$. All results are presented in nats.}
    \label{fig:ar_agn_capacities}
\end{figure}

We consider the MA-AWGN channel of order 1:
\begin{align}
    &Z_i = \alpha N_{i-1} + N_i \nonumber\\
    &Y_i = X_i + Z_i,
    \label{eq:MA_1_model}
\end{align}
where $N_i \sim\mathcal{N}(0, 1)$ are i.i.d., $X_i$ is the channel input sequence bound to the average power constraint $\mathbb
E\left[ X_i^2 \right] \leq P$, and $Y_i$ is the channel output. We consider both feedforward and feedback cases. 
The feedforward capacity can be calculated via the water-filling algorithm \cite{CovThom06}. 
When feedback is present, we consider the capacity characterization from \cite{yang2007feedback} as $-\log(x_0)$, where $x_0$ is a solution to a 4th order polynomial equation. In Figure \ref{fig:C_magn_comp}, we compare our DINE-based capacity estimator with the above solutions, again revealing clear correspondence.  

\subsection{MIMO Gaussian AR(1) Channel}
The AR(1) Gaussian channel is given by
\begin{align}
    &Z_i = \alpha Z_{i-1} + N_i \nonumber\\
    &Y_i = X_i + Z_i,
    \label{eq:ar_4d_model}
\end{align}
where $X_i\in\RR^4$ and $N_i\sim \cN(0,\mathrm{I}_4)$ where $\mathrm{I}_4$ is the 4-dimensional identity matrix.
We consider the power constraint $\mathrm{tr}(\mathrm{K}_{X_i})\leq P$ for some $P\in\RR_{\geq0}$, where $\mathrm{K}_X$ is the covariance matrix of $X$.
The feedforward capacity of \eqref{eq:ar_4d_model} is obtained by the water filling algorithm, considering both the spatial and frequency domains.
For the feedback capacity, the authors of \cite{sabag2021feedback} recently developed a method for calculating the capacity of a general class of MIMO Gaussian channels with memory through sequential convex optimization. This class subsumes the MIMO AR(1) channel as a special case. Figure \ref{fig:ar_4d} compares the performance of Algorithm \ref{alg:cap_est} with the above methods.
The convergence of the algorithm is shown in in Figure \ref{fig:ar_4d_convergence}, presenting a long evaluation over $10^5$ samples, taken every 20 training iterations. 
It is evident that our method converges in a relatively small number of iterations and the ground truth is attained in all considered cases.

\section{Concluding Remarks and Future Work}\label{sec:conclusion}
This work proposed a new neural estimation-optimization framework of the DI rate between two jointly stationary and ergodic stochastic processes.
Drawing upon recent neural estimation techniques and modifying the LSTM architecture, we developed the DINE, proved its consistency, and described its implementation.
Then, we utilized an auxiliary deep generative model for the input process to obtain a provably consistent joint estimation-optimization scheme of DI rate. The method enables estimating channel capacity when the channel model is unknown (but can be sampled) or when the optimization objective is not tractable, accounting for both feedback and feedforward scenarios. We provided an empirical study that validated our theory and demonstrated the accuracy of the proposed framework for capacity estimation of various channel examples. The capacity estimates demonstrated significant correspondence with known theoretical solutions and/or bounds, and the learned input model was shown to approximate capacity-achieving input distributions.


Our method enables consistent estimation of channel capacity without the typically imposed model assumptions. However, the obtained estimate generally does not lower or upper bound the true capacity value. In future work, we plan to explore modified neural estimation techniques that would give rise to such theoretical bounds. Another appealing avenue is utilizing the learned NDT-based input distribution, or an appropriate adaptation thereof, to obtain explicit capacity-achieving coding schemes. We also plan to extend our method to multiuser channels with arbitrary input and output spaces, targeting a unified and scalable framework of channel capacity estimation. Moreover, we will look to apply the proposed scheme to other time-series domains, such as control, computer vision, speech recognition, and reinforcement learning.

\section{Proofs}\label{sec:proofs}
\subsection{Proof of Theorem \ref{theorem:DINE_consistency}}\label{subsec:consistency_proof}
With some abuse of notation, let $\{(X_i,Y_i)\}_{i \in \ZZ}$ be the two-sided extension of the considered processes, and $\PP$ be the underlying stationary ergodic measure over $\sigma(\XX,\YY)$. An $n$-coordinate projection of $\PP$ is denoted by $P_{X^n Y^n}:=\mathbb{P}\big|_{\sigma(X^n,Y^n)}$, where $\sigma(X^n,Y^n)$ is the $\sigma$-algebra generated by $(X^n,Y^n)$. With this notation, $\Dn=(X^n,Y^n)\sim P_{X^n Y^n}$.
Lastly, let $\tilde{Y}\sim\mathsf{Unif}(\cY)$ (recall that $\cY\subset \RR^{d_y}$ is compact) be independent of $\{(X_i,Y_i)\}_{i \in \mathbb{Z}}$ and denote its distribution by $P_{\tilde{Y}}$. 
We divide the proof into three steps: variational representation, estimation from samples, and functional approximation. 

\underline{Representation of DI rate}.
We first write the DI rate as the limit of certain KL divergence terms. To do so, we use to following lemma:
\begin{lemma}[DI rate vs. $\boldsymbol{\DKL}$]\label{lemma:DI_rate_lim_DKL}
Let 
\begin{align*}
    \sD_{Y\|X}^{\infty}&:=\DKL\left(P_{Y^0_{-\infty}\| X^0_{-\infty}}\middle\| P_{Y^{-1}_{-\infty}\| X^{-1}_{-\infty}}\otimes P_{\widetilde{Y}}\middle|P_{X^{0}_{-\infty}\|Y^{-1}_{-\infty}}\right)\\
    \sD_{Y}^{\infty}&:=\DKL\left(P_{Y^0_{-\infty}}\middle\| P_{Y^{-1}_{-\infty}}\otimes P_{\widetilde{Y}}\right).
\end{align*}
Then we have
\begin{equation}\label{eq:DI_rate_DKL}
    \sI(\XX\to \YY) = \sD_{Y\|X}^{\infty}-\sD_{Y}^{\infty}.
\end{equation}
\end{lemma}
Lemma \ref{lemma:DI_rate_lim_DKL} is proven in Appendix \ref{appendix:DI_rate_lime_kl_proof}. The proof uses the stationarity of the considered processes and the monotone convergence theorem for the KL divergence (cf., e.g., \cite[Corollary 3.2]{polyanskiy2014lecture}).
We henceforth focus on estimating $\sD_{Y}^{\infty}$ and $\sD_{Y\|X}^{\infty}$.
Using the DV representation (Theorem \ref{theorem:DV}), we have 
\begin{subequations}
\begin{equation}
    \sD_{Y}^{\infty} = \sup_{f_y: \Omega_\cY \to \mathbb{R}}\mathbb{E}\left[ f_y\big(Y^0_{-\infty}\big) \right] -\log\mathbb{E}\left[ e^{f_y\big(Y^{-1}_{-\infty},\tilde{Y}\big)} \right]\label{eq:proof_dy_dv1},
\end{equation}
where $\Omega_\cY=\cY^0_{-\infty}$. For $\sD_{Y\|X}^{\infty}$, we use the KL divergence chain rule to write
\begin{equation*}
    \sD_{Y\|X}^{\infty} = \DKL\left(P_{X^{0}_{-\infty}\|Y^{-1}_{-\infty}}P_{Y^0_{-\infty}\| X^0_{-\infty}}\middle\| P_{X^{0}_{-\infty}\|Y^{-1}_{-\infty}}P_{Y^{-1}_{-\infty}\| X^{-1}_{-\infty}}\otimes P_{\widetilde{Y}}\right),
\end{equation*}
and via the DV theorem obtain
\begin{equation*}
    \sD_{Y\|X}^{\infty} = \sup_{f_{xy}: \Omega_{\cX\times\cY} \to \mathbb{R}}\mathbb{E}\left[ f_{xy}\big(X^0_{-\infty},Y^0_{-\infty}\big) \right] -\log\mathbb{E}\left[ e^{f_2\big(X^0_{-\infty},Y^{-1}_{-\infty},\tilde{Y}\big)}\right],\numberthis{}\label{eq:proof_dy_dv2}
\end{equation*}%
\end{subequations}
where $\Omega_{\cX\times\cY}=\cY^0_{-\infty}\times \cX^0_{-\infty}$.

We now provide a full treatment of \eqref{eq:proof_dy_dv1}. Afterwards, we refer back to \eqref{eq:proof_dy_dv2} and explain how its analysis reduces to that of \eqref{eq:proof_dy_dv1}, without repeating the argument.

\underline{Step 2: Estimation.}
The supremum in \eqref{eq:proof_dy_dv1} is achieved by
\begin{equation}\label{eq:t_star_seq}
    f^\star_{y,\infty} := \log\left(\frac{\dd P_{Y^0_{-\infty}}}{\dd(P_{Y^{-1}_{-\infty}}\otimes P_{\widetilde{Y}})}\right) \stackrel{(a)}{=} \log p_{Y_0|Y^{-1}_{-\infty}} -\log p_{\widetilde{Y}},
\end{equation}
where (a) holds because $P_{Y^0_{-\infty}} \ll P_{Y^{-1}_{-\infty}}\otimes P_{\widetilde{Y}}$ and both measures have Lebesgue densities. Since $\tilde Y$ is uniform, $p_{\widetilde{Y}}$ is a constant; denote $c_\cY:=\log\big(p_{\widetilde{Y}}(y)\big)$, for any $y\in\cY$.
We next show that the expectations in \eqref{eq:proof_dy_dv1} can be estimated with empirical means. Namely, for any $\epsilon>0$ and sufficiently large $n$, we have $\PP-$a.s. that
\begin{subequations}
\begin{align}
    &\hspace{0.9cm}\left| \mathbb{E}\Big[ f^\star_{y,\infty}\big(Y^{0}_{-\infty}\big)\Big] - \frac{1}{n}\sum_{i=0}^{n-1}f^\star_{y,i}\left(Y^0_{-i}\right) \right|< \frac{\epsilon}{8}\label{eq:d_y_convergence1}\\
    &\left|\log\left( \mathbb{E}\Big[ e^{f^\star_{y,\infty}\big(Y^{-1}_{-\infty},\widetilde{Y}\big)} \Big]\right) - \log\left(\frac{1}{n}\sum_{i=0}^{n-1}e^{f^\star_{y,i}\left(Y^{-1}_{-i},\widetilde{Y}\right)}\right) \right|< \frac{\epsilon}{8}\label{eq:d_y_convergence2},
\end{align}\label{eq:d_y_convergence}%
\end{subequations}
where $\{f^\star_{y,i}\}_{i\in\NN}$ is the sequence of supremum achieving elements of $\left\{\DKL\left(P_{Y^0_{-i}}\middle\| P_{Y^{-1}_{-i}}\otimes P_{\widetilde{Y}}\right)\right\}_{i\in\NN}$, with the $\PP-$a.s. limit $\lim_{i\to\infty}f^\star_{y,i} = f^\star_{y,\infty}$.
To simplify notation we denote the following empirical means over $n$ samples as
\begin{align}
    \EE_n[f^\star_y(Y^0_{-(n-1)})] &:= \frac{1}{n}\sum_{i=0}^{n-1}f^\star_{y,i}\left(Y^0_{-i}\right)\\
    \EE_n
    \left[e^{f^\star_y(\tilde{Y},Y^{-1}_{-(n-1)})}\right] &:= \frac{1}{n}\sum_{i=0}^{n-1}e^{f^\star_{y,i}\left(Y^{-1}_{-i},\widetilde{Y}\right)},\label{eq:En_def}
\end{align}
and invoke the generalized form of the asymptotic equipartition (AEP) theorem \cite{algoet1988sandwich}, as stated next. 

\begin{theorem}[Generalized AEP] \label{theorem_:general_aep}
Suppose $\mathbb{M}$ is a $v^{th}$ order Markov measure with a stationary transition kernel $\kappa(\dd X_v|X^{v-1}_0)$, and the finite-dimensional marginals of $\mathbb{M}$ are absolutely continuous w.r.t. the corresponding marginals of a stationary measure $\mathbb{P}$, i.e., if $\mathbb{P}$ is ergodic, $\EE$ is the expectation w.r.t. $\PP$ and $p_{X^0_{-(n-1)}}:= \frac{\dd\mathbb{P}}{\dd\mathbb{M}}\Bigr\rvert_{\sigma\big(X^0_{-(n-1)}\big)}$, then
\begin{align*}
    \frac{1}{n}\log\Big(p_{X^0_{-(n-1)}}(X_0,\dots,X_{-(n-1)})\Big) &= \frac{1}{n}\sum_{i=0}^{n-1} \log\Big(p_{X_0|X^{-1}_{-i}}\big(X_0 \big| X^{-1}_{-i}\big)\Big)\\
    &\hspace{-0.1cm}\xrightarrow[n \to \infty]{} \mathbb{E}\left[ \log p_{X_0|X^{-1}_{\infty}} (X_0|X_{-\infty}^{-1})\right],\qquad \PP-\mbox{\emph{a.s.}}\numberthis
\end{align*}
\end{theorem}
By Theorem \ref{theorem_:general_aep}, we obtain
\begin{equation}
    \lim_{n \to \infty}  \EE_n[f^\star_{y}(Y^0_{-(n-1)})] = \mathbb{E}\Big[ f^\star_{y,\infty}\big(Y^{0}_{-\infty}\big)\Big], \qquad \PP-\mbox{a.s.},
\end{equation}
where $f^\star_{y,i}:= \log p_{Y_0|Y^{-1}_{-i}} - c_\cY$.

Some additional work is needed to  justify \eqref{eq:d_y_convergence2}. First, by \cite[Proposition 2.6]{neveu1975discrete}, we have that the sequence $\Big(e^{f^\star_{y,n} + c_\cY}, \sigma\big(Y^{-1}_{-(n-1)},\widetilde{Y}\big)\Big) = \left(p_{Y_0|Y^{-1}_{-(n-1)}}, \sigma\big(Y^{-1}_{-(n-1)},\widetilde{Y}\big)\right)$ is a positive supermartingale converging a.s. to $p_{Y_0|Y^{-1}_{-\infty}}$, which equals $e^{f^\star_{y,\infty} + c_\cY}$ with $f^\star_{y,\infty}$ given in \eqref{eq:t_star_seq}. Consequently, $\{e^{f^\star_{y,n}}\}_{n=1}^\infty$ converges a.s. as a multiplication of the aforementioned sequence with a constant. 
We now apply a generalization of Birkhoff's ergodic theorem (due to Breiman
\cite[Theorem 1]{breiman1957individual}), as stated next.
\begin{theorem}[The generalized Birkhoff theorem]\label{theorem:modified_birkhoff}
Let $T$ be a metrically transitive $1-1$ measure preserving transformation\footnote{This translates into the condition $\PP(A) = \PP(T^{-1}(A))$ for any $A\in\cF$. We consider the time shift transformation.} of the probability space $(\Omega, \mathcal{F}, \mathbb{P})$  onto itself. Let $g_0(\omega), g_y(\omega), \dots$ be a sequence  of measurable functions on $\Omega$ converging a.s. to the function $g(\omega)$ such that $\mathbb{E}[\sup_k |g_k|] \leq \infty$. Then,
\begin{equation}
    \frac{1}{n}\sum_{k=1}^n g_k(T^k\omega) \xrightarrow{n \to \infty} \mathbb{E}[g], \qquad \PP-\mbox{a.s.}
\end{equation}
\end{theorem}
Applying Theorem \ref{theorem:modified_birkhoff} together with the continuous mapping theorem from \cite[Corollary 2]{mann1943stochastic}, 
we conclude that
\begin{equation*}
    \lim_{n \to \infty}\log\left(  \frac{1}{n}\sum_{i=0}^{n-1}e^{f^\star_{y,i}\big(Y^{-1}_{-i}\widetilde{Y}_{0}\big)}\right)= \log\left(   \mathbb{E}\left[ e^{f^\star_{y,\infty}\big(Y^{-1}_{-\infty},\widetilde{Y}\big)}\right]\right), \qquad \PP-\mbox{a.s.}\numberthis{}
\end{equation*}
This, in turn, implies \eqref{eq:d_y_convergence2} for a large enough $n$.

\underline{Step 3: Approximation.}
The last step is to approximate the functional space with the space of RNNs. namely, we define
\begin{equation}
    \hat{\sD}_Y(\Dn) := \sup_{g_y\in\GRNN^{Y}}\frac{1}{n}\sum_{i=0}^{n-1}g_y(Y^{0}_{-i}) - \log \left(\frac{1}{n}\sum_{i=0}^{n-1}e^{g_y(Y^{-1}_{-i},\widetilde{Y}_{0})}\right)\label{eq:cons_proof_dy_hat},
\end{equation}
and we want to show that for a given $\epsilon>0$, we know that
\begin{equation*}
    \left|\hat{\sD}_Y(\Dn)-\sD^{(\infty)}_Y\right|\leq\frac{\epsilon}{2}.
\end{equation*}
By Theorem \ref{theorem:DV}, we have
$$
\EE\left[f_{y,\infty}^\star(Y^0_{-\infty})\right] = \sD^{(\infty)}_Y,\quad \EE\left[f_{y,\infty}^\star(Y^{-1}_{-\infty}, \tilde{Y})\right] =1.
$$
We therefore bound the expression $\left|\hat{\sD}_Y(\Dn)-\EE\left[f_y^\star(Y^0_{-\infty})\right]\right|$. 
First, by the identity $
\log(x)\leq x-1$  for every $x\in\RR_{\geq0}$ we have
\begin{align*}
    &\Big{|}\hat{\sD}_Y(\Dn)-\EE\left[f_y^\star(Y^0_{-\infty})\right]\Big{|} \\
    &\hspace{0.8cm}=\left|\sup_{g_y\in\GRNN^{Y}}\frac{1}{n}\sum_{i=0}^{n-1}g_y(Y^{0}_{-i}) - \log \left(\frac{1}{n}\sum_{i=0}^{n-1}e^{g_y(\widetilde{Y},Y^{-1}_{-i})}\right)
    -\EE\left[f_y^\star(Y^0_{-\infty})\right]\right|\\
    &\hspace{0.8cm}\leq\left|\sup_{g_y\in\GRNN^{Y}}\frac{1}{n}\sum_{i=0}^{n-1}g_y(Y^{0}_{-i}) - \left(\frac{1}{n}\sum_{i=0}^{n-1}e^{g_y(\widetilde{Y}, Y^{-1}_{-i})}\right)
    +1-\EE\left[f_y^\star(Y^0_{-\infty})\right]\right|\\
     &\hspace{0.8cm}\leq\left|\sup_{g_y\in\GRNN^{Y}}\frac{1}{n}\sum_{i=0}^{n-1}g_y(Y^{0}_{-i}) - \left(\frac{1}{n}\sum_{i=0}^{n-1}e^{g_y(\widetilde{Y},Y^{-1}_{-i})}\right)
    +\EE\left[e^{f^\star_{y,\infty}(\widetilde{Y},Y^{-1}_{-\infty})}\right]-\EE\left[f_y^\star(Y^0_{-\infty})\right]\right|.
    \numberthis{}{}\label{eq:cons_proof_first_step}
\end{align*}
Due to \eqref{eq:d_y_convergence} and the a.s. convergence of $\{e^{f^\star_{y,n}}\}_{n\in\NN}$, there exists an integer $N\in\NN$ such that for every $n>N$
\begin{equation}
    \left|\EE_n\left[f_{y}^\star(Y^0_{-(n-1)})\right]-\EE\left[f_{y,\infty}^\star(Y^0_{-\infty})\right]\right|
\leq\frac{\epsilon}{8},\quad \left|\EE_n\left[e^{{f_{y}^\star}(\tilde{Y},Y^{-1}_{-(n-1)})}\right]-\EE\left[e^{f_{y,\infty}^\star(\tilde{Y},Y^{-1}_{-\infty})}\right]\right|
\leq\frac{\epsilon}{8}\label{eq:eps_est_bound}
\end{equation}
Plugging \eqref{eq:eps_est_bound} into \eqref{eq:cons_proof_first_step},we have
\begin{align*}
    \Big{|}\hat{\sD}_Y(\Dn)-\sD_Y^{\infty}\Big{|}&\leq \Bigg{|}\EE_n\left[e^{f_{y}^\star(\tilde{Y},Y^{-1}_{-(n-1)})}\right]-\EE_n\left[f_{y}^\star(Y^0_{-(n-1)})\right] \\
    &\hspace{1cm} -\sup_{g_y\in\GRNN^{Y}}\left\{\frac{1}{n}\sum_{i=0}^{n-1}g_y(Y^{0}_{-i}) - \left(\frac{1}{n}\sum_{i=0}^{n-1}e^{g_y(Y^{-1}_{-i},\widetilde{Y}_{0})}\right)\right\}\Bigg{|} + \frac{\epsilon}{4}.
\end{align*}
By assumption, $\{f^\star_{y,i}\}_{i\in\NN}$ is a sequence of functions convering a.s. to a function $f^{\star}_{y,\infty}$, uniformly bounded by some $M\in\RR_{\geq0}$. Since the exponent function is Lipschitz continuous with Lipschitz constant $e^M$ on the interval $(-\infty,M]$, we obtain
\begin{equation}
    \frac{1}{n}\sum_{i=1}^n e^{f_{y,i}^\star(\tilde{Y},Y^{-1}_{-i})}-e^{g_y(\tilde{Y},Y^{-1}_{-i})} \leq e^M\frac{1}{n} \sum_{i=1}^n \Big{|}f_{y,i}^\star(\tilde{Y},Y^{-1}_{-i})-g_y(\tilde{Y},Y^{-1}_{-i})\Big{|}.
\end{equation}
We conclude this stage by applying the universal approximation theorem for 
RNNs \cite{jin1995universal}.
To that end, we show that the sequence of supremum-achieving DV potentials are a dynamic system.
\begin{definition}[Dynamic system]\label{def:new_open_dyn_system}
 Let $d_{\mathsf{i}},d_{\mathsf{o}}, T \in \NN$, $\cZ\subseteq\RR^{d_{\mathsf{o}}}$ and $\cU\subseteq\RR^{d_{\mathsf{i}}}$ be open sets, $\cD_z\subseteq\cZ$ be a compact set and $f:\cZ\times\cU\mapsto\cZ$ be a continuous vector-valued function. Then, the system $\mathsf{Z}^{(d_{\mathsf{i}},d_{\mathsf{o}})}:=\{z_t\}_{t=1}^T$ defined by 
\begin{equation}
    z_{t+1} = f(z_t, u_t) 
\end{equation}
for $t\in\{1,\dots,T\}$ with some initial value $z_0\in\cD_z$
is a dynamic system.
\end{definition}
We define a $2$-dimensional system output $z_t = \left[p_{Y^0_{-(t-1)}}, \log p_{Y_0|Y^{-1}_{-(t-1)}}\right]$, where at each time step, the indices of the elements of $z_t$ are shifted back by a single step.
The system input is the new element $y_0$. 
Thanks to the universal approximation theorem for RNNs \cite[Theorem 2]{jin1995universal}, we can approximate this system by elements of the class $\GRNN^{Y}$ to arbitrary precision.
\begin{theorem}[Universal approximation for RNNs]\label{theorem:universal_approx_rnn}
Let $\epsilon>0$, $T\in\NN$, $\cU\subset\RR^{d_x}$ be an open set and $\mathsf{Z}^{(d_{\mathsf{i}},d_{\mathsf{o}})}$ be a dynamic system as in Definition \ref{def:new_open_dyn_system}. 
There exist a $k\in\NN$ and a $k$-neuron RNN $g\in\GRNN^{(d_{\mathsf{i}},d_{\mathsf{o}}, k)}$ (as in Definition \ref{def:RNN_function_class}), such that for any sequence of inputs $\{u_t\}_{t=1}^T\in\cU^T$, we have
\begin{equation}
    \max_{0\leq t \leq T} \| Z_t-g(U_t) \|_1\leq\epsilon.
\end{equation}
\end{theorem}
For given $\epsilon$, $M$, and $T=n$, denote by $g_y^\star\in\GRNN^{(d_{y},1, k)}$ the RNN 
such that the approximation error is uniformly bounded by $e^{-M}\frac{\epsilon}{4}$ for all $t=1\dots n$. Finally, we have
\begin{equation}
    \Big{|}\hat{\sD}_Y(\Dn)-\sD_Y^{(\infty)}\Big{|}\leq (1+e^M)\frac{1}{n} \sum_{i=1}^n \left|f_{y,i}^\star(\tilde{Y},Y^{-1}_{-i})-g^\star_y(\tilde{Y},Y^{-1}_{-i})\right|+\frac{\epsilon}{4}\leq \frac{\epsilon}{2}\label{eq:final_y}.
\end{equation}
This concludes the proof of \eqref{eq:proof_dy_dv1}.
For \eqref{eq:proof_dy_dv2}, note that
\begin{align*}
    f^\star_{xy,\infty} = \log\left(\frac{\mathrm{d}P_{X^{0}_{-\infty}\|Y^{-1}_{-\infty}}\otimes P_{Y^0_{-\infty}\| X^0_{-\infty}}}{\mathrm{d}P_{X^{0}_{-\infty}\|Y^{-1}_{-\infty}}\otimes P_{Y^{-1}_{-\infty}\| X^{-1}_{-\infty}}\otimes P_{\widetilde{Y}}}\right)
    = \log p_{Y_0|Y^{-1}_{-\infty}X^{0}_{-\infty}} -c_{\cY}
\end{align*}
achieves the supremum. Following similar arguments to those above, one may verify that
\begin{equation}\label{eq:final_xy}
    \left| \hat{\sD}_{Y\|X}(\Dn) - \sD_{Y\|X}^{\infty} \right| < \frac{\epsilon}{2}, \qquad \PP-\mbox{a.s.},
\end{equation}
where
\begin{equation*}
    \hat{\sD}_{Y\|X}(\Dn):= \sup_{g_{xy}\in\GRNN^{XY}}\frac{1}{n}\sum_{i=0}^{n-1}g_{xy}(Y^{0}_{-i},X^{0}_{-i})
    - \log \left(\frac{1}{n}\sum_{i=0}^{n-1}e^{g_{xy}(Y^{-1}_{-i},X^{0}_{-i},\widetilde{Y}_{0})}\right).\numberthis{} 
\end{equation*}
Combining \eqref{eq:final_y} and \eqref{eq:final_xy} concludes the proof.
$\hfill\square$

\subsection{Proof of Theorem \ref{theorem:ndt_consistency}}\label{sec:ndt_consistency_proof}
Let $\epsilon>0$ and $U^n\sim P^{\otimes n}_U$. 
Fix the USC $\{P_{Y_i|Y^{i-1}X^{i-1}}\}_{i\in\ZZ}$ as defined in Section \ref{subsec:ndt_theo}.
Recall that $\sX_s$ includes the class of stationary Markov processes of finite order and is therefore non-empty.
Thus, there exist some $\XX^{\epsilon}\in\sX_s$ such that $\left| \sI(\XX^\epsilon\to\YY)-\underline{\sC}_s \right|\leq\epsilon/3$ by its definition as a supremum over a non-empty set.
We denote a corresponding sample of $\XX^\epsilon$ and the channel by $\Dn^\epsilon=(X^{\epsilon,n},Y^{\epsilon,n})\sim\prod_{i=1}^nP_{X^\epsilon_{i}|X^{\epsilon,i-1}}P_{Y_i|X^{i}Y^{i-1}}$.
We have
\begin{align}
    \left|\underline{\sC}_s-\dine^\star(U^n)\right| &\leq \frac{\epsilon}{3} + \left|  \sI(\XX^\epsilon\to\YY)-\dine(\Dn^\epsilon)\right|+\Big| \dine(\Dn^\epsilon)-\dine^\star(U^n) \Big|  \nonumber\\
    &\leq \frac{2\epsilon}{3}+\left| \dine(\Dn^\epsilon)-\dine^\star(U^n) \right|\label{eq:ndt_step_dine_cons}\\
    &= \frac{2\epsilon}{3}+\inf_{\NDT\in\GRNN^X}\left|\hspace{1mm}\dine(\Dn^\epsilon)-\dine(\Dphi,\NDT) \right|\numberthis{}{}\label{eq:total_bound_dine_ndt},
\end{align} 
where \eqref{eq:ndt_step_dine_cons} follows from Theorem \ref{theorem:DINE_consistency} for a large enough $n\in\NN$, and $\dine(\Dn^\epsilon)$ is given in \eqref{eq:dine_est_def}.
Therefore, our goal is to bound the remaining term in \eqref{eq:total_bound_dine_ndt}, which quantifies the DINE error induced by using the approximating dataset $\Dphi$.

First, we show that the evolution of an RSP can be reformulated as an open dynamical system.
Namely, an open dynamical system with inputs $v^n$, states $s^n$ and outputs $x^n$ taking values in $\cV\subseteq\RR^{d_v}$, $\cS\subseteq\RR^{d_s}$, $\cX\subseteq\RR^{d_x}$, respectively, is given by following set of equations \cite[Eqn.~1]{schafer2006recurrent}.
\begin{subequations}
\begin{align}
    s_{t+1} &= f_1(s_t, v_t)\\
    x_t &= f_{xy}(s_t),
\end{align}\label{eq:open_dynamic_system}%
\end{subequations}
where $f_1$ is Borel measurable and $f_2\in\sC(\cS)$.
Recall that the evolution of $\XX\in\sX_\cS$ is described by the relation 
\begin{subequations}
\begin{align}
    &S_i = f_\mathsf{s}(X_i, S_{i-1})\\
    &P_{X_i|X^{i-1}, S_{i-1}} = P_{X_i|S_{i-1}}.\label{eq:rsp_x_markov_state}
\end{align}\label{eq:rsp_eqns}%
\end{subequations}
To show that \eqref{eq:rsp_eqns} adheres to the relation presented in \eqref{eq:open_dynamic_system}, we utilize the following lemma.
\begin{lemma}[Functional representation of RSPs]\label{lemma:rsp_func_rep}
For any $\XX\in\sX_{\cS}$ with state process $\BS$ and an i.i.d. process $\WW$ with $W_1\sim P_W\in\lebmeas(\cW)$ and $\cW\subseteq\RR^{d_x}$, there exists a function $f_\mathsf{x}:\cS\times\cW\to\cX$ such that
\begin{equation}
    X_i=f_\mathsf{x}(S_{i-1}, W_{i}),\qquad \forall i\in\NN.
\end{equation}
\end{lemma}
The proof is given in Appendix \ref{sec:rsp_func_rep_proof}. It follows from the stationarity of $\XX$, the FRL and Lemma~\ref{lemma:gen_inverse_transform}.
Lemma \ref{lemma:rsp_func_rep} provides us with $f_{\mathsf{x}}$ such that
$$
    S_i = f_\mathsf{s}(X^\epsilon_i, S_{i-1}), \quad
    X^\epsilon_i=f_\mathsf{x}(S_{i-1}, U_{i}).
$$
As a final step towards the relation \eqref{eq:open_dynamic_system},
denote $\tilde{S}_i := (S_i,U_i)$ and $V_i := (U_i,X^\epsilon_i)$ and define $\tilde{f}_{\mathsf{s}}$ such that the first $d_s$ components of $\tilde{S_i}$ are calculated from $f_{\mathsf{s}}(S_{i-1},X^\epsilon_{i-1})$ and the rest of its components comprise of replacing $U_{i-1}$ with $U_i$.
We therefore have the following open-dynamical system representation.
\begin{align*}
    \tilde{S}_{i} &= \tilde{f}_{\mathsf{s}}(\tilde{S}_{i-1},V_i)\\
    X^\epsilon_i &= f_\mathsf{x}(\tilde{S}_{i})\numberthis{}{}\label{eq:rsp_proof_eq}.
\end{align*}

Having an open-dynamical system representation of $\XX^\epsilon$, we will approximate it with RNNs, due to the following Theorem \cite[Theorem~2]{schafer2006recurrent}.
\begin{theorem}[Universal approximation of open dynamical systems]
Let $n\in\NN$, $\epsilon>0$, and let $u_t\in\RR^{d_{\mathsf{i}}}$, $s_t\in\RR^{d_s}$ and $x_t\in\RR^{d_{\mathsf{o}}}$ be the inputs, states and outputs of an open dynamical system for $t=1,\dots,n$.
Then,
there exists $k\in\NN$ and  $\NDT\in\GRNN^{(d_{\mathsf{i}},d_{\mathsf{o}},k)}$ such that
\begin{equation}
    \max_{t=1,\dots,n}\left\| \NDT(u^i)-x_i \right\|_1\leq\epsilon.
\end{equation}
\end{theorem}
Therefore, take $\epsilon'>0$ and fix sample $u^n\in\cU^n$ drawn according to $P^{\otimes n}_U$; there exists $k\in\NN$ and $\NDT\in\GRNN^X$ such that 
\begin{equation}
    \max_{t=1\dots n} \left\| x^\epsilon_i(u^i) - x^\phi_i(u^i) \right\|_1\leq \epsilon'.
\end{equation}

Our next step is to bound $\big\|y^{\epsilon}_i - y^{\phi}_i \big\|_1$ in terms of $\big\|x^{\epsilon}_i - x^{\phi}_i \big\|_1$ for $i=1,\dots,n$.
To that end, consider the following lemma.
\begin{lemma}\label{lemma:channel_error_bound}
Let $T\in\NN$ and $\YY$ be the output of the USC described in Section \ref{subsec:dine_ndt_consistency} with $f_\mathsf{y}$ and $f_{\mathsf{z}}$ satisfying Assumption \ref{assumption:dine_ndt_channel_assumption} with Lipschitz constants $M_y$ and $M_z$, respectively.
Then, for any $n\in\NN$, every pair of input sequences $(x^{1,n},x^{2,n})$ such that
$
    \max_{t=1,\dots,n}\|x^1_t-x^2_t\|_1\leq\eta
$, we have
\begin{equation*}
    \max_{i=1,\dots,T}\|y^1_t-y^2_t\|_1\leq\frac{M_y (2-M_z(M_y+1))}{1 - M_z(M_y+1)}\eta.
\end{equation*}
\end{lemma}
The proof of Lemma \ref{lemma:channel_error_bound} is in Appendix \ref{appendix:channel_error_bound_proof}.
We further denote $\alpha(M_y,M_z) :=\frac{M_y (2-M_z(M_y+1))}{1 - M_z(M_y+1)}$.

Finally, we have
\begin{align*}
    \left|\hspace{1mm}\dine(\Dn^\epsilon)-\dine(\Dphi,\NDT) \right|
    &\leq\frac{1}{n}\sum_{i=1}^n\left|  g_y(y^\epsilon_i|y^{\epsilon,i-1})-g_y(y^\phi_i|y^{\phi,i-1}) \right|\\
    &\hspace{1.7cm}+\frac{e^M}{n}\sum_{i=1}^n\left| g_y(\tilde{y}|y^{\epsilon, i-1}) - g_y(\tilde{y}|y^{\phi,i-1}) \right|\\
    &\hspace{1.7cm} + \frac{1}{n}\sum_{i=1}^n\left|  g_{xy}(y_i^\epsilon|y^{\epsilon,i-1},x^{\epsilon,i})-g_{xy}(y_i^\phi|y^{\phi,i-1},x^{\phi,i}) \right|\\
    &\hspace{1.7cm}+\frac{e^M}{n}\sum_{i=1}^n\left| g_{xy}(\tilde{y}|y^{\epsilon,i-1},x^{\epsilon,i}) - g_{xy}(\tilde{y}|y^{\phi,i-1},x^{\phi,i}) \right| \numberthis{}{}\label{eq:ndt_proof_appprox_error}.
\end{align*}
By assumption, $g_y$ and $g_{xy}$ are Lipschitz continuous with Lipschitz constants $M_1$, $M_2$, respectively.
Consequently, we have
\begin{align*}
    &\left|\hspace{1mm}\dine(\Dn^\epsilon)-\dine(\Dphi,\NDT) \right|\\
    &\qquad\qquad\qquad\leq\frac{(M_1+M_2)(1+ e^M)}{n}\sum_{i=1}^n\big\|  y^\epsilon_i-y^\phi_i \big\|_1 + \frac{M_2(1+ e^M)}{n}\sum_{i=1}^n\big\|  x_i^\epsilon-x_i^\phi \big\|_1\\
    &\qquad\qquad\qquad\leq \left((M_1+M_2)(1+ e^M)\alpha(M_y,M_z) + M_2(1+ e^M)\right)\epsilon'\numberthis{}{} \label{eq:ndt_conv_dine_bound}.
\end{align*}
Take a large enough $k\in\NN$ such that \eqref{eq:ndt_conv_dine_bound} is bounded by $\epsilon/3$. 
As the above steps hold for any realization of $U^n$ and $K^n$, the inequality \eqref{eq:ndt_conv_dine_bound} holds $\PP-a.s.$
This concludes the proof. $\hfill\square$

\begin{remark}[Lipschitz assumption]
Lemma \ref{lemma:channel_error_bound} calls for Assumption \ref{assumption:dine_ndt_channel_assumption} due to the recursive nature of the proposed channel, i.e., $Y_i$ and $Z_i$ indirectly depend on their past values and the induced error accumulates over time.
By restricting $f_{\mathsf{z}}$ to be a function of only $X_i$, the resulting Lipschitz constants $M_z$, $M_y$ are no longer bound to $M_y(M_z+1)<1$.
\end{remark}

\begin{remark}[Channels with feedback]
To account for the feedback scenario, we first consider a \textit{conditional} version of $\sX_{\cS}$ that allows conditioning on past channel outputs. The state $S_i$ is then taken as a function of $(X_i,S_{i-1},Y_{i-1})$ and we require $P_{X_i|X^{i-1},Y{i-1},S^{i-1}} = P_{X_i|S_i}$. Lemma \ref{lemma:rsp_func_rep} follows immediately, as the FRL holds even when conditioning on additional random variables.
The rest of the proof follows by adding $Y_i$ to the $i$th input of $f_{\mathsf{s}}$.
\end{remark}

\subsection{Proof of Theorem \ref{theorem:opt_ndt}}\label{subsec:proof_ndt_strcut}
Let $U\sim P_U$ and $P_{Y|X}$ be a given transition kernel.
Throughout this proof we employ the tools of Gaussian smoothing developed in \cite{nietert21} (see also 
\cite{goldfeld2019,Goldfeld2020GOT,goldfeld2020asymptotic,sadhu2021,goldfeld2022limit}). 
To this end, we denote the isotropic $d_x$-dimensional Gaussian distribution with $\cN_\sigma:=\cN(0,\sigma^2\mathrm{I}_{d_x})$ with the corresponding PDF $\varphi_\sigma$.
Let $P_{X^\star}$ be the MI maximizing input distribution for $P_{Y|X}$ and denote its corresponding smoothed distribution with $P_{X_\sigma^\star}:=P_{X^\star}*\cN_\sigma$.
For any choice of $\sigma>0$ we have $P_{X_\sigma^\star}\in\lebmeas(\cX)$, which implies the existence of the bijection $T_{X_\sigma^\star}\in\sC^1(\cU,\cX)$ due to Lemma  \ref{lemma:gen_inverse_transform}.
We utilize the universal approximation theorem for NNs with arbitrary finite output dimension \cite[Corollary~1]{schafer2006recurrent}.
\begin{lemma}[Universal approximation of NNs]\label{lemma:uni_approx_nn_output}
Let $\sC(\cX,\cY)$ be the class continuous functions $f:\cX\to\cY$ where $\cX\subset\RR^{\mathsf{d}_i}$ is compact and $\cY\subseteq\RR^{\mathsf{d}_o}$.
Then, the class of NNs $\GNN^{(\mathsf{d}_i,\mathsf{d}_o)}$ is dense in $\sC(\cX,\cY)$, i.e., for every $f\in\sC(\cX,\cY)$ and $\epsilon>0$, there exist $g\in\GNN^{(\mathsf{d}_i,\mathsf{d}_o)}$ such that
$
\|f-g\|_\infty\leq\epsilon.
$
\end{lemma}
By Lemma \ref{lemma:uni_approx_nn_output}, we can construct a sequence of functions $\{\NDTk\}_{k\in\NN}\subset\GNN^{(d_x,d_x)}$ such that $\|\NDTk-T_{X_\sigma^\star}^{-1}\|_\infty\to 0$. Setting $P_{X^{\phi_k}}:=\NDTkpush P_U$, we therefore obtain $P_{X^{\phi_k}}\rightharpoonup P_{X_\sigma^\star}$, where $\rightharpoonup$ denotes weak convergence of probability measures.\footnote{a sequence of measures $\{\mu_n\}_{n\in\NN}$ converges weakly to a measure $\mu$ if $\int f \dd\mu_n \to \int f \dd\mu$ for any continuous and bounded function $f$.}
As a consequence of the weak convergence, the compactness of $\cU$, and the continuity of $\NDTk$, we have convergence of second moments, i.e., $\int_{\RR^{d_x}}\|x\|^2\dd P_{X^{\phi_k}}(x)\to\int_{\RR^{d_x}}\|x\|^2\dd P_{X_\sigma^\star}(x)$.
As weak convergence plus convergence in $2$-th moments is equivalent to convergence under the $2$-Wasserstein distance, we obtain  $\sW_2(P_{X^{\phi_k}},P_{X_\sigma^\star})\to 0$ as $\sigma\to0$\footnote{In general, we have convergence of any $p$th moment for any $p<\infty$, therefore, convergence of $p$th Wasserstein distance.} (cf., e.g., \cite[Theorem~7.12]{villani2021topics}).

Given a non-increasing sequence $\sigma_i \searrow 0$, it is readily verified that  $P_{X_{\sigma_i}^\star}{\rightharpoonup}P_{X^\star}$ and the second moments converge as well. Indeed, the former follows because weak convergence is equivalent to pointwise convergence of characteristic functions together with the fact that the characteristic function of $\cN_\sigma$ never vanishes; the latter follows from a uniform integrability argument.
We therefore have $\sW_2(P_{X_{\sigma_i}^\star},P_{X^\star})\stackrel{i\to\infty}{\longrightarrow}0$.
To bound $\sW_2(P_{X^\star},P_{X^{\phi_k}})$ we perform two steps of approximation; first, we approximate $P_{X^\star}$ with $P_{X_{\sigma_i}^\star}$ which is then approximated with $P_{X^{\phi_k}}$.
Take large enough $i,k\in\NN$ such that the corresponding $2$-Wasserstein metrics are bounded by $\epsilon/2$ and apply the triangle inequality to result with
\begin{equation}
    \sW_2(P_{X^\star},P_{X^{\phi_{k}}}) \leq \sW_2(P_{X^\star}, P_{X_{\sigma_i}^\star})  + \sW_2(P_{X_{\sigma_i}^\star},P_{X^{\phi_{k}}})\leq \epsilon.\label{eq:wasser_eps}
\end{equation}
We stress that $k$ is taken w.r.t. the chosen index of $\sigma_i$, but omit this in our notation for simplification.

All considered input-output pairs are distributed with the fixed transition kernel $P_{Y|X}$, therefore, they only differ by the input distribution.
To bound the difference \eqref{eq:ndt_struct_mine_bound}, we consider two intermediate steps of approximation.
First, we consider the MI induced by the approximation of $X^\star$ by an element from the sequence of its Gaussian smoothed counterpart for some $\sigma_i$, denoted $X^\star_{\sigma_i}:=X^\star+Z_{\sigma_i}$, where $Z_{\sigma_i}\sim \cN(0,\sigma_i^2\mathrm{I}_{d_x})$.
Then, our task is to approximate the MI induced by $X^\star_{\sigma_i}$ with the MI induced by $X^{\phi_{k}}:=h_{\phi,k}(U)$.
To do so, we apply an intermediate step of an approximation of both elements with a smoothed version of $X^{\phi_{k}}$, denoted $X^{\phi_{k}}_{\sigma_\ell}:=h_{\phi,k}(U)+Z_{\sigma_\ell}$, where $Z_{\sigma_\ell}\sim \cN(0,\sigma_\ell^2\mathrm{I}_{d_x})$ for some $\sigma_\ell$.
The last step consists of approximating the MI induced by $X^{\phi_{k}}$ and its $n$-sample MINE approximation calculated from $\Dphik=\{(h_{\phi,k}(U_i), Y_i)\}_{i=1}^n$.
By the triangle inequality, we have
\begin{align*}
    \left|\sC - \mine\big(\Dphik\big) \right| &\leq
    \left|\sI(X^\star;Y^\star) - \sI(X_{\sigma_i}^\star;Y_{\sigma_i}^\star) \right|
    + \left|\sI(X_{\sigma_i}^\star;Y_{\sigma_i}^\star) - \sI(X^{\phi_{k}}_{\sigma_\ell};Y^{\phi_{k}}_{\sigma_\ell}) \right| \\
    &\hspace{1.2cm}+ \left|\sI(X^{\phi_{k}}_{\sigma_\ell};Y^{\phi_{k}}_{\sigma_\ell}) - \sI(X^{\phi_{k}};Y^{\phi_{k}}) \right|
    + \left|\sI(X^{\phi_{k}};Y^{\phi_{k}}) - \mine\big(\Dphik\big) \right|.\numberthis{}{}\label{eq:opt_ndt_triangle}
\end{align*}
To bound the first term in  \eqref{eq:opt_ndt_triangle}, we utilize the weak lower semicontinuity of MI \cite[Section~3.5.2]{polyanskiy2014lecture}, i.e, $P_{X^\star_{\sigma_i},Y^\star_{\sigma_i}}\rightharpoonup P_{X^\star,Y^\star}$ implies
\begin{equation}
    \sI(X^\star;Y^\star)\leq\liminf_{i\to\infty}\sI(X^\star_{\sigma_i};Y^\star_{\sigma_i}).\label{eq:lsc}
\end{equation}
With some abuse of notation, extract a subsequence $(X^\star_{\sigma_j},Y^\star_{\sigma_j})_{j\in\NN}$ that achieves the RHS of \eqref{eq:lsc}.
Recall that $P_{X^\star_j, Y^\star_j}\rightharpoonup P_{X^\star,Y^\star}$.
Along with the weak lower semicontinuity of MI and the fact that $X^\star$ achieves capacity for the fixed $P_{Y|X}$, there exist $j\in\NN$ such that 
\begin{equation}
    |\sC - \sI(X^\star_{\sigma_j};Y^\star_{\sigma_j})| \leq \frac{\epsilon}{3}.
\end{equation}



To bound the second term in \eqref{eq:opt_ndt_triangle}, we consider a non-increasing sequence $\sigma_\ell \searrow 0$ and denote $P_{X^{\phi_k}_{\sigma_\ell}} :=P_{X^{\phi_k}}*\cN_{\sigma_l}$, where $k_j\in\NN$ is taken such that the bound \eqref{eq:wasser_eps} still holds.
The second term in \eqref{eq:opt_ndt_triangle} can then be bounded as follows.
\begin{align*}
    \left|\sI(X_{\sigma_j}^\star;Y_{\sigma_j}^\star) - \sI(X^{\phi_k}_{\sigma_\ell};Y^{\phi_k}_{\sigma_\ell}) \right| &= \left| \DKL(P_{X_{\sigma_j}^\star Y_{\sigma_j}^\star}\|P_{X_{\sigma_j}^\star}P_{Y_{\sigma_j}^\star}) - \DKL(P_{X^{\phi_k}_{\sigma_\ell}Y^{\phi_k}_{\sigma_\ell}}\|P_{X^{\phi_k}_{\sigma_\ell}}P_{Y^{\phi_k}_{\sigma_\ell}}) \right|\\
    &= \left|\EE_{P_{X_{\sigma_j}^\star Y_{\sigma_j}^\star}}\left[ \log\frac{p_{X_{\sigma_j}^\star Y_{\sigma_j}^\star}}{p_{X^{\phi_k}_{\sigma_\ell}Y^{\phi_k}_{\sigma_\ell}}} \right] 
    + \EE_{P_{X^{\phi_k}_{\sigma_\ell}Y^{\phi_k}_{\sigma_\ell}}}\left[ \log\frac{p_{X^{\phi_k}_{\sigma_\ell}}p_{Y^{\phi_k}_{\sigma_\ell}}}{p_{X_{\sigma_j}^\star}p_{Y_{\sigma_j}^\star}} \right]\right|\\
    &= \left|\DKL(P_{X_{\sigma_j}^\star}\|P_{X^{\phi_k}_{\sigma_\ell}})+ \EE_{P_{X^{\phi_k}_{\sigma_\ell}Y^{\phi_k}_{\sigma_\ell}}}\left[ \log\frac{p_{X^{\phi_k}_{\sigma_\ell}}p_{Y^{\phi_k}_{\sigma_\ell}}}{p_{X_{\sigma_j}^\star}p_{Y_{\sigma_j}^\star}}\right] \right|\numberthis{}\label{eq:dkl_bound_mi_dif_two_terms}
\end{align*}
where \eqref{eq:dkl_bound_mi_dif_two_terms} follows from the construction of both joint distributions with the same transition kernel $P_{Y|X}$.
The second term in \eqref{eq:dkl_bound_mi_dif_two_terms} can be represented as follows.
\begin{align*}
    \EE_{P_{X^{\phi_k}_{\sigma_\ell}Y^{\phi_k}_{\sigma_\ell}}}\left[ \log\frac{p_{X^{\phi_k}_{\sigma_\ell}}p_{Y^{\phi_k}_{\sigma_\ell}}}{p_{X_{\sigma_j}^\star}p_{Y_{\sigma_j}^\star}} \right] &= \int_\cX \int_\cY \log\frac{p_{X^{\phi_k}_{\sigma_\ell}}(x)p_{Y^{\phi_k}_{\sigma_\ell}}(y)}{p_{X_{\sigma_j}^\star}(x)p_{Y_{\sigma_j}^\star}(y)}p_{X^{\phi_k}_{\sigma_\ell}Y^{\phi_k}_{\sigma_\ell}}(x,y)\dd x \dd y\\
    &= \int_{\cX}\log\frac{p_{X^{\phi_k}_{\sigma_\ell}}(x)}{p_{X_{\sigma_j}^\star}(x)}p_{X^{\phi_k}_{\sigma_\ell}}(x)\dd x + \int_\cY \log\frac{p_{Y^{\phi_k}_{\sigma_\ell}}(y)}{p_{Y_{\sigma_j}^\star}(y)}p_{Y^{\phi_k}_{\sigma_\ell}}(y) \dd y\\
    &= \DKL(P_{X^{\phi_k}_{\sigma_\ell}}\|P_{X_{\sigma_j}^\star}) + \DKL(P_{Y^{\phi_k}_{\sigma_\ell}}\|P_{Y_{\sigma_j}^\star})\numberthis{}{}\label{eq:dkl_bound_mi_diff_2}.
\end{align*}
Plug \eqref{eq:dkl_bound_mi_diff_2} into \eqref{eq:dkl_bound_mi_dif_two_terms} and apply the data-processing inequality for KL divergences to obtain
\begin{equation}
     \left|\sI(X_{\sigma_j}^\star;Y_{\sigma_j}^\star) - \sI(X^{\phi_k}_{\sigma_\ell};Y^{\phi_k}_{\sigma_\ell}) \right| \leq 2\left( \DKL(P_{X_{\sigma_j}^\star}\|P_{X^{\phi_k}_{\sigma_\ell}}) + \DKL(P_{X^{\phi_k}_{\sigma_\ell}}\|P_{X_{\sigma_j}^\star}) \right).\label{eq:kl_bound_on_mi_diff}
\end{equation}
We note that both KL terms are well defined as both $P_{X_{\sigma_j}^\star}$ and $P_{X^{\phi_k}_{\sigma_\ell}}$ are defined and positive over the same space as Gaussian smoothed distributions.
We will now upper bound the RHS of \eqref{eq:kl_bound_on_mi_diff} with $\sW_2(P_{X_{\sigma_j}^\star},P_{X^{\phi_k}_{\sigma_\ell}})$, using the following theorem \cite[Proposition~1]{polyanskiy2016wasserstein}.
\begin{theorem}
Let $U$ and $V$ be random vectors with finite second moments. If both $U$ and $V$ are $(c_1,c_2)$-regular, then 
\begin{align}
    \DKL(P_U\|P_V) + \DKL(P_V\|P_U) \leq 2\Delta,
\end{align}
where $P_U$ is $(c_1,c_2)$-regular if
\begin{equation}
    \|\nabla\log p_U(u)\|_2\leq c_1\|u\|_2+c_2,
\end{equation}
and
\begin{equation}
    \Delta := \left(\frac{c_1}{2}\Big(\sqrt{\EE\left[\|V\|_2^2\right]}+\sqrt{\EE\left[\|U\|_2^2\right]}\Big)+c_2 \right)\sW_2(P_U,P_V).
\end{equation}
\end{theorem}
The $(c_1,c_2)$ regularity of $P_{X_{\sigma_j}^\star}$ and $P_{X^{\phi_k}_{\sigma_\ell}}$ follows from the Gaussian smoothing of $P_{X^\star}$ and $P_{X^{\phi_k}}$ such that the regularity parameters depend on $\sigma_j$ and $\sigma_\ell$ \cite[Proposition~2]{polyanskiy2016wasserstein}.
Note that $P_{X^{\phi_k}}\in\cP_2(\cX)$ follows from the compactness of $h_{\phi,k}(\cU)$.
Consequently, $P_{X^{\star}_{\sigma_j}}\in\cP_2(\cX)$ as $\EE\big[\|X^\star_{\sigma_j}\|^2\big]=\EE\big[\|X^\star\|^2\big]+\big[\|Z_{\sigma_j}\|^2\big]$, both having finite second moment.
We can therefore bound \eqref{eq:kl_bound_on_mi_diff} with $\sW_2(P_{X^\star},P_{X^{\phi_k}_{\sigma_\ell}})$, which by the triangle inequality, amounts to
\begin{equation}
\sW_2(P_{X^\star_{\sigma_j}},P_{X^{\phi_k}_{\sigma_\ell}}) \leq \sW_2(P_{X^\star_{\sigma_j}},P_{X^{\phi_k}}) + \sW_2(P_{X^{\phi_k}},P_{X^{\phi_k}_{\sigma_\ell}}).\label{eq:wasser_traingle_smoothed}
\end{equation}
For given $\epsilon$ take $k$ and $\ell$ large enough and utilize the weak continuity of $\sW_2(P_{X^\star_{\sigma_j}},P_{X^{\phi_k}_{\sigma_\ell}})$ to obtain an $\epsilon/6$ bound on \eqref{eq:kl_bound_on_mi_diff}.

We now describe the bound of the third term in the RHS of \eqref{eq:opt_ndt_triangle}.
First, represent each MI as a combination of differential entropies to obtain the following bound
\begin{align*}
    \left|\sI(X^{\phi_k}_{\sigma_\ell};Y^{\phi_k}_{\sigma_\ell}) - \sI(X^{\phi_k};Y^{\phi_k}) \right| &\leq \left|\sh(X^{\phi_k}_{\sigma_l})-\sh(X^{\phi_k})\right| + \left|\sh(Y^{\phi_k}_{\sigma_l})-\sh(Y^{\phi_k})\right| \\
    &\hspace{4.5cm}+ \left|\sh(X^{\phi_k}_{\sigma_l},Y^{\phi_k}_{\sigma_l})-\sh(X^{\phi_k},Y^{\phi_k})\right|.\numberthis{}{}\label{eq:mi_ent_decomp}
\end{align*}
We will utilize the following Theorem \cite[Theorem~1]{godavarti2004convergence}.
\begin{theorem}[Convergence of differential entropies]\label{theorem:diff_ent_cont}
Let $(X_i)_{i\in\NN}$ be a sequence of continuous random variables with PDFs $(f_i)_{i\in\NN}$ and $X$ be a continuous random variable with PDF $f$ such that $f_i\to f$ pointwise. If 
\begin{subequations}
\begin{equation}
    \max\left\{\|f_i\|_{\infty}, \|f\|_{\infty}\right\}\leq A_1<\infty\label{eq:cond_1_ent}
\end{equation}
\begin{equation}
    \max\left\{\int\|x\|^\kappa f_i(x) \dd x,\int\|x\|^\kappa f(x) \dd x\right\} \leq A_2 < \infty\label{eq:cond_2_ent},
\end{equation}%
\end{subequations}
for some $\kappa>1$ and for all $i\in\NN$, then $h(X_i)\to h(X)$.
\end{theorem}

We will now show that the conditions of Theorem \ref{theorem:diff_ent_cont} hold in our case, focusing on $\kappa=2$.
Note that if such conditions hold for the input and output distributions, they hold for the joint distribution as well.
To justify the pointwise convergence of PDFs we introduce the notion of asymptotic equicontinuity (a.e.c.).
A function $f$ is a.e.c. on $x\in\cX$ if for every $\epsilon>0$ there exist $\delta(x,\epsilon)$ and $n_0(x,\epsilon)$ such that whenever $\|x-y\|_1<\delta(x,\epsilon)$, then $|f_n(x_1)-f_n(x_2)|<\epsilon$ for any $n>n_0$.
We use following theorem \cite[Theorem~1]{sweeting1986converse}.
\begin{theorem}
Let $(P_n)_{n\in\NN}\subset\cP(\cX)$ with PDFs $(p_n)_{n\in\NN}$. The following statements are equivalent.
\begin{enumerate}
    \item $(p_n)_{n\in\NN}$ are a.e.c. on $\cX$ and $P_n\rightharpoonup P$.
    \item $p_n\to p$ pointwise, where $p$ is the continuous PDF of $P$.
\end{enumerate}
\end{theorem}
Recall that both $P_{X^{\phi_k}_{\sigma_\ell}}$ and $P_{Y^{\phi_k}_{\sigma_\ell}}$ weakly converge to $P_{X^{\phi_k}}$ and $P_{Y^{\phi_k}}$, respectively.
The a.e.c. property of $p_{x^{\phi_k}_{\sigma_\ell}}$ follows from its structure is a convolution with a Gaussian density, as follows
\begin{align}
   \left|p_{X^{\phi_k}_{\sigma_\ell}}(x_1)-p_{X^{\phi_k}_{\sigma_\ell}}(x_2)\right| &= \left|\int_{\RR^{d_x}}p_{X^{\phi_k}}(x_1-u)\varphi_{\sigma_\ell}(u)\dd u - \int_{\cX}p_{X^{\phi_k}}(x_2-u)\varphi_{\sigma_\ell}(u)\dd u\right|\\
   &= \left|\int_{\RR^{d_x}}\varphi_{\sigma_\ell}(u)\left(p_{X^{\phi_k}}(x_1-u) - p_{X^{\phi_k}}(x_2-u)\right)\dd u\right|\\
   &<\left|\int_{\RR^{d_x}}\varphi_{\sigma_\ell}(u)\epsilon\dd u\right|\label{eq:fx_cont}\\ 
   &=\epsilon,
\end{align}
where \eqref{eq:fx_cont} follows from the continuity of $p_{X^{\phi_k}_{\sigma_\ell}}$, taking appropriate $\delta>0$.
The a.e.c. property of $p_{Y^{\phi_k}_{\sigma_\ell}}$ follows from the same steps and the continuity of $p_{Y|X}$ on $\cY$.
The boundedness of $p_{X^\phi_k}$ follows from the extreme value theorem, as it is a continuous function on $h_{\phi_k}(\cU)$.
The PDF $p_{X^{\phi_k}_{\sigma_\ell}}$ is integrable due to Fubini's theorem.
Consequently, the PDFs $p_{X^{\phi_k}_{\sigma_\ell}}$, $p_{Y^{\phi_k}}$ and $p_{Y^{\phi_k}_{\sigma_\ell}}$ are bounded as they are continuous integrable PDFs on $\RR^{d_x}$.
The second moment of ${X^{\phi_k}}$ is bounded by the compactness of $h_{\phi,k}(\cU)$ and the second moment bound of $X^{\phi_k}_{\sigma_\ell}$ follows from
$$
\EE\left[\|X^{\phi_k}_{\sigma_\ell}\|_2^2\right]= \EE\left[\|X^{\phi_k}\|_2^2\right] + \EE\left[\|Z_{\sigma_\ell}\ |_2^2\right]= \EE\left[\|X^{\phi_k}\|_2^2\right] + d_x\sigma_\ell^2<\infty ,
$$
where $Z_{\sigma_\ell}\sim\cN(0,\sigma_\ell\mathrm{I}_{d_x})$ is independent of $X^{\phi_k}$.
The second moment bound for $Y^{\phi_k}$ and $Y^{\phi_k}_{\sigma_\ell}$ follows from the assumption on $P_{Y|X}$.
We can therefore apply Theorem \ref{theorem:diff_ent_cont} to bound the differences of differential entropies in \eqref{eq:mi_ent_decomp}.
Take $\ell$ large enough such that both \eqref{eq:mi_ent_decomp} and \eqref{eq:kl_bound_on_mi_diff} are bounded by $\epsilon/6$.


Finally, the fourth term in \eqref{eq:opt_ndt_triangle} can be bounded by $\epsilon/3$ for large enough $n\in\NN$ using the MINE consistency \cite[Theorem~2]{belghazi2018mutual}, which concludes the proof. $\hfill\square$

\section{Proofs of Lemmas}

\subsection{Proof of Lemma \ref{lemma:gen_inverse_transform}}\label{proof:gen_inverse_transform_proof}
Let $X\sim P_X\in\lebmeas(\cX)$ with $\cX\subseteq \RR^{d_x}$, and let $(x_1,\dots,x_{d_x})$ be an arbitrary ordering of the elements of its realization $x$.
Let $T_X$ as defined in \eqref{eq:d_cdf_map} and denote its output with $Z:=T_X(X_1,\dots,X_{d_x})$ such that $Z_i=\big[T_X(X)\big]_i$.
First, following the steps of the proof of \cite[Proposition~3]{huang2018neural}, we know that $Z\sim\mathsf{unif}[0,1]^{d_x}$ due to \cite[Theorem 1]{hyvarinen1999nonlinear}, providing us with part 1.

To show that $T_X$ is a bijection, first, note that for $i=1,\dots,d_x$
$
    z_i =  \PP(X_i\leq x_i| x^{i-1}) 
$.
Following the steps of the proof of \cite[Proposition~2]{huang2018neural},
denote $F_i:=F_{X_i|X^{i-1}}$ and the Jacobian martix of $T_X$ with $\mathrm{J}_{T_X}$.
We have
\begin{equation}
    \mathrm{J}_{T_X} := \begin{pmatrix}
\frac{\partial F_1}{\partial x_1} & \frac{\partial F_1}{\partial x_2} & \cdots & \frac{\partial F_1}{\partial x_{d_x}} \\
\frac{\partial F_2}{\partial x_1} & \frac{\partial F_2}{\partial x_2} & \cdots & \frac{\partial F_2}{\partial x_{d_x}} \\
\vdots  & \vdots  & \ddots & \vdots  \\
\frac{\partial F_{d_x}}{\partial x_1} & \frac{\partial F_{d_x}}{\partial x_2} & \cdots & \frac{\partial F_{d_x}}{\partial x_{d_x}} 
\end{pmatrix} 
=
\begin{pmatrix}
\frac{\partial F_1}{\partial x_1} & 0 & \cdots & 0 \\
\frac{\partial F_2}{\partial x_1} & \frac{\partial F_2}{\partial x_2} & 0 \cdots & 0 \\
\vdots  & \vdots  & \ddots & \vdots  \\
\frac{\partial F_{d_x}}{\partial x_1} & \frac{\partial F_{d_x}}{\partial x_2} & \cdots & \frac{\partial F_{d_x}}{\partial x_{d_x}}
\end{pmatrix}, 
\end{equation}
i.e., $\mathrm{J}_{T_X}$ is a lower triangular matrix and its determinant is therefore given by a product of conditional PDFs, which is strictly positive for any $x$ in the interior of $\cX$.
Therefore, $T_X$ is a bijection almost everywhere with its inverse $T_X^{-1}$, defined similar to \eqref{eq:d_cdf_map}, i.e.,
$$
    T_X^{-1}(T_X(x^i), z^{i-1}) = x_i.
$$
Finally, for $U:=(U_1,\dots,U_{d_x})\sim\mathsf{Unif}[0,1]^d$ we have $T_X^{-1}(U) = X$, following the proof of  \cite[Proposition~2]{huang2018neural}. $\hfill\square$

\subsection{Proof of Lemma \ref{lemma:DI_rate_lim_DKL}}\label{appendix:DI_rate_lime_kl_proof}
Recall that
\begin{align*}
    D_{Y\|X}^{n}&:=\DKL\left(P_{Y^n\| X^n}\middle\| P_{Y^{n-1}\| X^{n-1}}\otimes P_{\widetilde{Y}}\middle|P_{X^n\|Y^{n-1}}\right)\\
    D_{Y}^{n}&:=\DKL\left(P_{Y^n}\middle\| P_{Y^{n-1}}\otimes P_{\widetilde{Y}}\right).
\end{align*}
We first show that $\sI(\XX\to\YY)= \lim_{n \to \infty}\left(D_{Y\|X}^{n} - D_{Y}^{n}\right)$.
Recall that (see Section \ref{subsec:di})
\begin{equation}\label{eq:proof_di_ent}
     \sI(X^n\to Y^n)=\sh(Y^n)-\sh(Y^n\|X^n),
\end{equation}
and expand 
\begin{subequations}
\begin{align}
    \sh(Y^n)&=\sh_{\mathsf{CE}}\left(P_{Y^n},P_{Y^{n-1}}\otimes P_{\widetilde{Y}}\right) -\DKL\left(P_{Y^n}\middle\| P_{Y^{n-1}}\otimes P_{\widetilde{Y}}\right),\label{eq:hy_hce_dkl}\\
    \sh(Y^n \| X^n)&= \sh_{\mathsf{CE}}\left(P_{Y^n \| X^n},P_{Y^{n-1} \| X^{n-1}}\otimes P_{\widetilde{Y}}\middle|P_{X^n\|Y^{n-1}}\right)\nonumber\\
    &\hspace{3cm}-\DKL\left(P_{Y^n\| X^n}\middle\| P_{Y^{n-1}\| X^{n-1}}\otimes P_{\widetilde{Y}}\middle|P_{X^n\|Y^{n-1}}\right).\label{eq:hxy_hce_dkl}
\end{align}%
\end{subequations}
Subtraction yields
\begin{align*}\label{eq:Entropy_decomposition_proof}
    \sI(X^n \to Y^n) &= \Big{(}\sh_{\mathsf{CE}}(P_{Y^n},P_{Y^{n-1}}\otimes P_{\widetilde{Y}}) \\
    &\hspace{2.5cm}- \sh_{\mathsf{CE}}(P_{Y^n \| X^n},P_{Y^{n-1} \| X^{n-1}}\otimes P_{\widetilde{Y}}|P_{X^{0}_{-(n-1)}\|Y^{-1}_{-(n-1)}})\Big{)}\\
    &+ \Big{(}\DKL(P_{Y^n\| X^n}\| P_{Y^{n-1}\| X^{n-1}}\otimes P_{\widetilde{Y}}|P_{X^{0}_{-(n-1)}\|Y^{-1}_{-(n-1)}}) \\
    &\hspace{4.5cm}- \DKL(P_{Y^n}\| P_{Y^{n-1}}\otimes P_{\widetilde{Y}})\Big{)}.\numberthis
\end{align*}

Denote the residual cross-entropy terms by $\sh_{\mathsf{CE},Y}$ and $\sh_{\mathsf{CE},Y \| X}$, respectively. By stationarity and since $\widetilde{Y}\indep\XX$, we further obtain
\begin{align*}
    \sh_{\mathsf{CE},Y} - \sh_{\mathsf{CE},Y \| X} &= \mathbb{E}\left[ -\log P_{Y^{-1}_{-(n-1)}}\otimes P_{\widetilde{Y}} (\widetilde{Y},Y^{-1}_{-n}) \right] \\
    & \hspace{2.5cm}- \mathbb{E}\left[ -\log P_{Y^{-1}_{-(n-1)} \| X^{-1}_{-(n-1)}}\otimes P_{\widetilde{Y}}(\widetilde{Y},Y^{-1}_{-n}) \right]\\
    &= \mathbb{E}\left[ -\log P_{Y^{-1}_{-(n-1)}} (Y^{-1}_{-n}) \right] - \mathbb{E}\left[ -\log P_{Y^{-1}_{-(n-1)} \| X^{-1}_{-(n-1)}} (Y^{-1}_{-n}) \right]\\
    &\hspace{2.55cm}+ \mathbb{E}\left[ -\log P_{\widetilde{Y}} (\widetilde{Y})\right] - \mathbb{E}\left[ -\log P_{\widetilde{Y}} (\widetilde{Y}) \right]\\
    &= \sh(Y^{-1}_{-(n-1)}) - \sh(Y^{-1}_{-(n-1)}\|X^{-1}_{-(n-1)})\\
    &= \sI(X^{n-1} \to Y^{n-1}),
\end{align*}
Plugging this term into \eqref{eq:Entropy_decomposition_proof} implies
\begin{align*}
    D_{Y\|X}^{n} - D_{Y}^{n} &= \sI(X^{n-1} \to Y^{n}) - \sI(X^{n} \to Y^{n-1})\\
   &= \sI(X^0_{-(n-1)};Y_0|Y^{-1}_{-(n-1)})\\
    &= \sh(Y_0|Y^{-1}_{-(n-1)}) - \sh(Y_0| Y^{-1}_{-(n-1)}X^0_{-(n-1)}) \numberthis{}\label{eq:ent_diff}.
\end{align*}
We now use the following theorem, restated from  \cite[Theorem 4.2.1]{CovThom06}.

\begin{theorem}[Entropy rate of stationary processes]\label{lemma:ent_rate_station}
For a stationary process $\{ Y_n \}_{n \in \mathbb{Z}}$, the following limits exist and are equal:
\begin{equation}
    \lim_{n \to \infty}{\frac{1}{n}\sh(Y^0_{-(n-1)})} = \lim_{n \to \infty}{\sh(Y_0|Y^1_{-(n-1)})}.
\end{equation}
\end{theorem}
Together with \eqref{eq:ent_diff}, the lemma implies
\begin{align*}
    \lim_{n \to \infty}{D_{Y\|X}^{n} - D_{Y}^{n}} &= \lim_{n \to \infty}\sh(Y_0|Y^{-1}_{-(n-1)}) - \sh(Y_0| Y^{-1}_{-(n-1)}X^0_{-(n-1)})\\
    &= \lim_{n \to \infty}\frac{1}{n}\Big(\sh\left(Y^0_{-(n-1)}\right) - \sh\left(Y^0_{-(n-1)} \| X^0_{-(n-1)}\right)\Big)\\
    &= \sI(\XX \to \YY).
\end{align*}

Our last step is to identify the limiting KL divergence terms using the monotone convergence theorem (cf., e.g., \cite[Corollary 3.2]{polyanskiy2014lecture}).
\begin{theorem}[$\DKL$ monotone convergence]\label{theorem:MCT}
The following holds:
\begin{align*}\label{eq:eps_1}
    D_{Y}^{n} &\nearrow \DKL\left(P_{Y^0_{-\infty}}\middle\| P_{Y^{-1}_{-\infty}}\otimes P_{\widetilde{Y}}\right)\\
    D_{Y\|X}^{n} &\nearrow \DKL\left(P_{Y^0_{-\infty}\|X^0_{-\infty}}\middle\| P_{Y^{-1}_{-\infty}\| X^{-1}_{-\infty}}\otimes P_{\widetilde{Y}}\middle|P_{X^0_{-\infty}}\right)\numberthis.
\end{align*}
\end{theorem}
Recalling the definition of 
$D_{Y}^{\infty}$ and $D_{Y\|X}^{\infty}$, this concludes the proof. \hfill\(\square\)

\subsection{Proof of Lemma \ref{lemma:rsp_func_rep}}\label{sec:rsp_func_rep_proof}
Let $\XX\in\sX_{\cS}$ with corresponding stationary state process $\BS$. 
By joint stationarity we have $P_{X_n|S_n}=P_{X|S}$ for any $n\in\ZZ$.
To construct the desired relation we utilize the FRL \cite[Theorem 1]{li2018strong}.
\begin{theorem}[Functional representation lemma]\label{lemma:func_rep}
For any pair of random variables $(X,Y)\sim P_{XY}$ (over a Polish space with a Borel probability measure) with $\sI(X;Y)<\infty$, there exists a random variable $Z$ independent of $X$ such that $Y$ can be expressed as a function $g(X,Z)$.
\end{theorem}
By Theorem \ref{lemma:func_rep} we know that there exist a random variable $V\sim P_V$ and a function $f_\mathsf{x}$ such that
\begin{equation}\label{eq:strong_func_rep}
    X_n = f_\mathsf{x}(V_n,S_n).
\end{equation}
As $P_{X_n|S_n}$ is independent of $n$, \eqref{eq:strong_func_rep} holds for any $n$ with the same choice of $f_\mathsf{x}$ and time-invariant distribution on $V_n$, i.e., define a sequence $\{V_n\}_{n\in\ZZ}\stackrel{i.i.d}{\sim}P_V$, we have
\begin{equation}\label{eq:strong_func_rep_seq}
    X_n = f_\mathsf{x}(V_n,S_n). 
\end{equation}
Let $U\sim\mathsf{Unif}[0,1]^{d_x}$ and $T_V$ be as defined in \ref{subsec:opt_ndt}. By Lemma \ref{lemma:gen_inverse_transform},
$
    V = T_V^{-1}(U)
$
for $U\sim\mathsf{Unif}([0,1]^{d_x})$.
Take $W\sim P_W$ and let $T_W$ be as in \ref{subsec:opt_ndt}.
Lemma \ref{lemma:gen_inverse_transform} shows that $T_W\sim\mathsf{Unif}[0,1]^d$.
We therefore construct the composite function
$\tilde{f}_\mathsf{v} := T_V^{-1} \circ T_W:\cW\mapsto\cV$.
By construction, $V=\tilde{f}_\mathsf{v}(W)$.
Plugging $\tilde{f}_{\mathsf{v}}$ into \eqref{eq:strong_func_rep_seq}, we have
$$
    X_i = f_\mathsf{x}(S_{n-1},\tilde{f}_{\mathsf{v}}(W_i)),
$$
which completes the proof. $\hfill\square$

\subsection{Proof of Lemma \ref{lemma:channel_error_bound}}\label{appendix:channel_error_bound_proof}
Let $\eta>0$, fix $i\in\{1,2,\dots,n\}$ and let $x^{1,n}$, $x^{2,n}$ and $k^n$ be realizations of $X^{1,n}$, $X^{2,n}$ and $K^n$, respectively.
Let $y^{j,n}$ and $z^{j,n}$ be generated according to $x^{j,n}$ and $k^n$ for $j=1,2$.
Let  $\Delta_{x,i}, \Delta_{z,i}$ and $\Delta_{y,i}$ be the $L^1$ distance of the channel inputs, states and outputs at the $i$th step, e.g., $\Delta_{x,i} = \|x^1_i-x^2_i\|_1$.
By the Lipschitz property of $f_{\mathsf{y}}$ and $f_{\mathsf{z}}$ and the triangle inequality, we have
\begin{subequations}
\begin{align}
    \Delta_{y,i} &\leq M_y\left\|(x_i^1,z_i^1,k_i)-(x_i^2,z_i^2,k_i)\right\|_1\leq M_y\big(\Delta_{x,i} + \Delta_{z,i}\big)\label{eq:ucsc_proof_y_bound}\\
    \Delta_{z,i} &\leq M_y\left\|(x_i^1,y_i^1,z^1_{i-1})-(x_i^2,y_i^2,z^2_{i-1})\right\|_1\leq M_z\big(\Delta_{x,i} + \Delta_{y,i} + \Delta_{z,i-1}\big)\label{eq:delta_z}.
\end{align}%
\end{subequations}
Combining \eqref{eq:ucsc_proof_y_bound} and \eqref{eq:delta_z}, we obtain
\begin{equation}
    \Delta_{z,i} \leq (M_z + M_zM_y)\Delta_{x,i} + (M_z + M_zM_y)\Delta_{z,i-1}. \label{eq:ucsc_recursive_eq}
\end{equation}
Recursively applying \eqref{eq:ucsc_recursive_eq} yields
\begin{equation}
    \Delta_{z,i} \leq \sum_{j=0}^{i-1}(M_z + M_zM_y)^j\Delta_{x,i-j}\label{eq:finite_sum_i}.
\end{equation}
Upper bound \eqref{eq:finite_sum_i} with the infinite sum and assume $\max_{i=1,\dots,n}\Delta_{x,i}\leq\eta$. We have
\begin{equation}
    \Delta_{z,i} \leq \eta\sum_{j=0}^{\infty}(M_z + M_zM_y)^j = \eta \frac{1}{1-M_z(M_y+1)}\label{eq:ucsc_proof_inf_sum},
\end{equation}
where the sum converges due to Assumption \ref{assumption:dine_ndt_channel_assumption}.
Plug \eqref{eq:ucsc_proof_inf_sum} into \eqref{eq:ucsc_proof_y_bound} to obtain
\begin{equation}
    \Delta_{y,i} \leq \eta\left( \frac{M_y(2-M_z(M_y+1))}{1-M_z(M_y+1)} \right),\label{eq:usc_bound_proof_final_eq}
\end{equation}
which holds for any $i\leq n$.
The inequality \eqref{eq:usc_bound_proof_final_eq} holds for any realization of $P_K^{\otimes n}$. $\hfill\square$

\newpage
\bibliographystyle{unsrt}
\bibliography{ref.bib}
\end{document}

%% file: Figures/dine_tikz.tex
\usetikzlibrary{arrows.meta, positioning, calc}

\tikzstyle{block_D} = [rectangle, rounded corners, minimum width=2.5cm, minimum height=1.4cm, align=center, draw=black, fill= black!5]

\tikzstyle{block_E} = [rectangle, rounded corners, minimum width=0.8cm, minimum height=0.8cm, align=center, draw=black, fill= black!5]

\tikzstyle{block_C} = [rectangle, minimum width=1cm, minimum height=0.01cm, draw=white, fill=white]
 
\tikzstyle{small_b} = [rectangle, minimum width=0.000001cm, minimum height=0.000001cm, draw=none, fill=black!20]

\begin{tikzpicture}[]

\node (enc) [small_b] at (-1,0) {};

\node [block_D, right=3cm of enc] (sampler)  {\large RNN\\ \large $\theta_y$};

\draw [-{Latex[width=3mm]}] (enc) -- node[above=1mm] {\large $Y_i$} (sampler);
 
\node [block_D, below right = 1.3cm and 0cm of enc] (reference)  {\large Sampler $P_{\tilde{Y}}$};

\node (k) [inner sep=0,minimum size=0,right=1.38cm of enc] {};

\draw [-{Latex[width=3mm]}] (k) --  (reference.north);

\draw[-{Latex[width=3mm]}] (reference.east) -- node[below right = 0.1cm and 0.1cm]{\large $\tilde{Y}_i$} +(1,0) -|  (sampler.south);

\node (channel) [block_D, right=4cm of sampler] {\large Loss\\ \large $\widehat{\mathsf{D}}_Y(\Dn,g_{\theta_y})$};

\draw [-{Implies},double,-{Latex[width=3mm]}] (sampler) -- node[above=1mm] { \large$g_{\theta_y}(Y_i|Y^{i-1})$} (channel);
\draw [-{Implies},double,-{Latex[width=3mm]}] (sampler) -- node[below=1mm] { \large$g_{\theta_y}(\tilde{Y}_i|Y^{i-1})$} (channel);


\draw [-{Implies},double,-{Latex[width=3mm]}] (channel.north) -- node[above left=0.3cm and 1.1cm] {\large $\nabla_{\theta_y}\widehat{\mathsf{D}}_Y(D_n,g_{\theta_y})$} +(0,0.5) -| ([yshift = 0.0cm]sampler.north);

\end{tikzpicture}

%% file: Figures/modified_lstm.tex
\begin{tikzpicture}[]


\node (lstm_t) [lstm_block] at (-2,0) {};

\node (fl_block_tilde_t) [small_block, below right = -3cm and -3.6cm of lstm_t] {\Large$f_{\mathsf{L}}$};

\node (fl_block_t) [small_block, below right = -3cm and -1.72cm of lstm_t] {\Large$f_{\mathsf{L}}$};

\node (inv_input_s_t_m_1) [inv1_block, below left = 0.5cm and 3cm of fl_block_t] {\Large$S_{t-1}$};

\node (inv_input_y_t) [inv1_block, below right = 2.5cm and -0.74cm of fl_block_t] {\Large$Y_{t}$};

\node (inv_input_y_t_tilde) [inv1_block, below right = 2.4cm and -0.74cm of fl_block_tilde_t] {\Large$\tilde{Y}_{t}$};

\node (inv_out_s_t) [inv1_block, above = 1.75cm  of fl_block_t] {\Large$S_{t}$};

\node (inv_out_s_tilde_t) [inv1_block, above = 1.75cm  of fl_block_tilde_t] {\Large$\tilde{S}_{t}$};

\node (lstm_t_next) [lstm_block, right = 2cm of lstm_t] {};

\node (fl_block_tilde_t_next) [small_block, below right = -3cm and -3.6cm of lstm_t_next] {\Large$f_{\mathsf{L}}$};

\node (fl_block_t_next) [small_block, below right = -3cm and -1.72cm of lstm_t_next] {\Large$f_{\mathsf{L}}$};


\node (inv_input_y_t_next) [inv1_block, below right = 2.5cm and -0.99cm of fl_block_t_next] {\Large$Y_{t+1}$};

\node (inv_input_y_t_next_tilde) [inv1_block, below right = 2.4cm and -0.99cm of fl_block_tilde_t_next] {\Large$\tilde{Y}_{t+1}$};

\node (inv_out_s_t_next) [inv1_block, above = 1.75cm  of fl_block_t_next] {\Large$S_{t+1}$};

\node (inv_out_s_tilde_t_next) [inv1_block, above = 1.75cm  of fl_block_tilde_t_next] {\Large$\tilde{S}_{t+1}$};


\draw [-{Latex[width=3mm]}] (inv_input_y_t) --  ([xshift = 0.3cm]fl_block_t.south);

\draw [-{Latex[width=3mm]}] (inv_input_y_t_tilde) --  ([xshift = 0.3cm]fl_block_tilde_t.south);

\draw [-{Latex[width=3mm]}] (fl_block_t) --  (inv_out_s_t);

\draw [-{Latex[width=3mm]}] (fl_block_tilde_t) --  (inv_out_s_tilde_t);

\draw [-{Latex[width=3mm]}] (inv_input_s_t_m_1) -|  ([xshift = -0.3cm]fl_block_tilde_t.south);

\draw [-{Latex[width=3mm]}] (inv_input_s_t_m_1)  -| (-2.7,-1.19) arc[start angle=180, end angle=0,radius=0.07cm] ([xshift = -0.3cm]fl_block_t.south);

\draw [-{Latex[width=3mm]}] (-2.57,-1.21)  -| ([xshift = -0.3cm]fl_block_t.south);


\draw [-{Latex[width=3mm]}] (inv_input_y_t_next) --  ([xshift = 0.3cm]fl_block_t_next.south);

\draw [-{Latex[width=3mm]}] (inv_input_y_t_next_tilde) --  ([xshift = 0.3cm]fl_block_tilde_t_next.south);

\draw [-{Latex[width=3mm]}] (fl_block_t_next) --  (inv_out_s_t_next);

\draw [-{Latex[width=3mm]}] (fl_block_tilde_t_next) --  (inv_out_s_tilde_t_next);


\draw [-{Latex[width=3mm]}] (-1.07,2.2) -| (0.8,-1.05) -- (2.6,-1.05) -| ([xshift = -0.3cm]fl_block_tilde_t_next.south) ;

\draw [-{Latex[width=3mm]}] (-1.07,2.2) -| (0.8,-1.05)  -- (3.31,-1.05) arc[start angle=180, end angle=0,radius=0.07cm] -- (4.6,-1.05) -| ([xshift = -0.3cm]fl_block_t_next.south) ;

\draw [-{Latex[width=3mm]}] (4.95,2.2) -| (6.7,-1.05) |- (7.3,-1.05);

\end{tikzpicture}

%% file: Figures/ndt_tikz.tex
\usetikzlibrary{arrows.meta, positioning, calc}

\tikzstyle{block_D} = [rectangle, rounded corners, minimum width=1.5cm, minimum height=1.6cm, align=center, draw=black, fill= black!5]

\tikzstyle{block_E} = [rectangle, rounded corners, minimum width=0.8cm, minimum height=0.8cm, align=center, draw=black, fill= black!5]

\tikzstyle{block_C} = [rectangle, minimum width=1cm, minimum height=0.01cm, draw=white, fill=white]
 
\tikzstyle{small_b} = [rectangle, minimum width=0.000001cm, minimum height=0.000001cm, draw=none, fill=black!20]

\begin{tikzpicture}[]

\node (input_u) [inner sep=0,minimum size=0] at (-2,0) {};

\node (input_y) [inner sep=0,minimum size=0, below right = 1cm and 0.8cm of input_u] {};

\node (inv_y_1) [inner sep=0,minimum size=0, left = 0.5cm of input_y] {};
\node (inv_y_2) [inner sep=0,minimum size=0, above = 0.5cm of input_y] {};
\node (inv_y_3) [inner sep=0,minimum size=0, left = 0.3cm of inv_y_1] {};
\draw (inv_y_1) --  ([yshift = 0.5cm]inv_y_2);
\draw (inv_y_3) --  (inv_y_1);

\node [block_D, below right =-0.3cm and 1.8cm of input_u] (rnn)  {\large RNN};

\draw [-{Latex[width=3mm]}] (input_u.east) --  node[above right = 1mm and 0.05cm] {\Large $U_i$} ([yshift = 0.5cm]rnn.west);

\draw [-{Latex[width=3mm]}] (input_y) -- node[above =1mm] {\Large $Y_i$} ([yshift = -0.5cm]rnn.west);

\node [block_D, right =0.5cm of rnn] (fcn)  {\large FCN};
\draw [-{Latex[width=3mm]}] (rnn.east) --  (fcn.west);

\node [block_D, right =0.5cm of fcn] (cons)  {\large Constraint};
\draw [-{Latex[width=3mm]}] (fcn.east) --  (cons.west);

\node (inv) [inner sep=0,minimum size=0,right=0.5cm of cons] {};
\node (inv_1) [inner sep=0,minimum size=0,right=0.7cm of inv] {};

\draw  (cons.east) --  (inv.west);
\draw [-{Latex[width=3mm]}] (inv) -- node[above right = 1mm and 0mm] {\Large $X_i$} (inv_1);

\node (delta) [block_E, above left = 1.2cm and 1cm of inv] {\Large$\Delta$};
\node (inv_2) [inner sep=0,minimum size=0,above=0.8cm of inv] {};
\node (inv_3) [inner sep=0,minimum size=0,above=1.6cm of inv] {};
\node (inv_4) [inner sep=0,minimum size=0,above=0.8cm of rnn] {};

\draw (inv_3) --  (delta.east);
\draw (inv) --  (inv_3);
\draw (delta.west) --  (inv_4);
\draw [-{Latex[width=3mm]}] (inv_4) --  (rnn.north);

\end{tikzpicture}

%% file: Figures/DINE_NDT.tex

\usetikzlibrary{arrows.meta, positioning, calc}

\tikzstyle{block_D} = [rectangle, rounded corners, minimum width=2.5cm, minimum height=1.4cm, align=center, draw=black, fill= black!5]

\tikzstyle{block_E} = [rectangle, rounded corners, minimum width=0.8cm, minimum height=0.8cm, align=center, draw=black, fill= black!5]

\tikzstyle{block_C} = [rectangle, minimum width=1cm, minimum height=0.01cm, draw=white, fill=white]
 
\tikzstyle{small_b} = [rectangle, minimum width=0.000001cm, minimum height=0.000001cm, draw=none, fill=black!20]


\begin{tikzpicture}[]
\node (enc) [block_D] at (-1,0) {\Large NDT\\\Large $\phi$};
\node (channel) [block_D, right=4.5cm of enc] {\Large Channel\\  $P(Y_t|X^{t-1},Y^{t-1})$};
\node (DINE) [block_D, below right =1cm and 0.5cm of enc] {\large DINE  \\ \large$\theta_y,\theta_{xy}$ };

\node (k) [inner sep=0,minimum size=0,left=1cm of channel] {}; 
\node (k2) [inner sep=0,minimum size=0, below left =0.4cm and 1cm of channel] {}; 
\node (k3) [inner sep=0,minimum size=0, below right =0.6cm and 1.6cm of channel] {}; 
\node (k_noise) [inner sep=0,minimum size=0,left=1cm of enc] {}; 

\draw [-{Latex[width=3mm]}] (enc) -- node[above=1mm] {\Large $X^\phi_t$} (channel);

\draw [-{Latex[width=3mm]}] (k_noise) -- node[above=1mm] {\Large $U_t$} (enc);

\node [small_b, right=1.5cm of channel] (invisible) {};
\draw (channel) -- node[above=1mm] {\Large $Y^\phi_t$} (invisible);

\draw [-{Latex[width=3mm]}] (k2) -- +(0,0) -| ([xshift = 0.75cm]enc);
\draw [-{Implies},double,-{Latex[width=3mm]}] (DINE) --node[below =1mm] {\hspace{-5cm}\large $\nabla_{\phi} \widehat{\mathsf{I}}_{\mathsf{DI}}(D_n^\phi,g_{\theta_y},g_{\theta_{xy}},h_\phi)$} +(-1.3,0) -| ([xshift = -2cm]enc);

\draw [-{Latex[width=3mm]}] (k) -- +(0,-1.7) |- ([yshift = 0.3cm]DINE.east);
\draw[-{Latex[width=3mm]}] (invisible) -- +(0,-1.9) |- ([yshift = -0.3cm]DINE.east);

\draw[-{Latex[width=3mm]}] (invisible) -- +(0,1.5) -| (enc);
\node (FB_space) [block_C, above left =0.68cm and 1cm of channel] {};
\node (FB_end) [block_C, above right =0.5cm and -1.7cm of FB_space] {};
\node (FB_delta) [block_E, left = 1.2cm of FB_space] {\Large$\Delta$};
\draw (FB_space.east) -- (FB_end);

\node (DINE_space) [block_C, left =0.7cm of DINE] {};
\node (DINE_end) [block_C, above right =0.4cm and -1.6cm of DINE_space] {};
\draw [-{Implies},double](DINE_space.east) -- (DINE_end);

\node (X_delta) [block_E, below left = -0cm and 2.4cm of channel] {\Large$\Delta$};

\end{tikzpicture}
